\newcommand{\bea}{\begin{eqnarray}}
\newcommand{\ena}{\end{eqnarray}}
\newcommand{\vs}[1]{\vspace{#1 mm}}
\newcommand{\hs}[1]{\hspace{#1 mm}}
\renewcommand{\a}{\alpha}
\renewcommand{\b}{\beta}
\renewcommand{\c}{\gamma}
\renewcommand{\d}{\delta}
\newcommand{\e}{\epsilon}
\newcommand{\s}{\sigma}

\newcommand{\dsl}{\pa \kern-0.5em /}

\newcommand{\pa}{\partial}
\newcommand{\vp}{\varphi}
\newcommand{\nn}{\nonumber\\}
\newcommand{\p}[1]{(\ref{#1})}

\newcommand{\gsim}{\, \mbox{\raisebox{-1.ex}
{$\stackrel{\textstyle>}{\textstyle\sim}$}}\,}
\newcommand{\lsim}{\, \mbox{\raisebox{-1.ex}
{$\stackrel{\textstyle<}{\textstyle\sim}$}}\,}

\documentclass{ws-ijmpa}
\usepackage{graphicx,amssymb}

\begin{document}

\markboth{Nobuyoshi Ohta}
{Accelerating Cosmologies and Inflation from M/Superstring Theories}

\catchline{}{}{}

\title{ACCELERATING COSMOLOGIES AND INFLATION FROM M/SUPERSTRING THEORIES}

\author{NOBUYOSHI OHTA}

\address{Department of Physics, Osaka University,
Toyonaka, Osaka 560-0043, Japan \\
ohta@phys.sci.osaka-u.ac.jp}

\maketitle

\pub{Preprint OU-HET 500}{Invited review to be published in IJMPA
}

\begin{abstract}
We review the recent developments in obtaining accelerating cosmologies
and/or inflation from higher-dimensional gravitational theories,
in particular superstring theories in ten dimensions and M-theory in
eleven dimensions.
We first discuss why it is difficult to obtain inflationary behavior
in the effective low-energy theories of superstring/M-theory, i. e.
supergravity theories. We then summarize interesting solutions including
S-branes that give rise to accelerating cosmologies and inflationary
solutions in M-theory with higher order corrections.
Other approaches to inflation in the string context are also briefly
discussed.

\keywords{Accelerating Cosmologies; Inflation; M/Superstring theories.}
\end{abstract}

\section{Introduction}

Our universe is believed to have undergone an inflationary evolution
in the early epoch. Why is this so? The reason is that this mechanism
naturally explains the following major questions in the current
cosmology:\cite{ksato,guth}
\begin{itemize}
\item
Horizon problem:
Why is the early universe so highly homogeneous
beyond causally connected regions?
\item
Flatness problem:
Why does the present universe appear so extremely flat?
\end{itemize}

In fact the recent cosmological observation seems to confirm its
existence.\cite{WMAP} Not only that, it is discovered that the present
universe still expands with acceleration (late-time acceleration).
It is not difficult to construct cosmological models with these
features in the Einstein gravity by suitable modifications.
For example, if one introduces a positive cosmological constant,
it is possible to get de Sitter expansion of exponential type.
A scalar field with positive potential can also be used for this purpose.
Also it has been pointed out that higher order curvature terms can
do similar job.\cite{staro}

However it is desirable to
derive such a model from the fundamental theories of particle physics
that incorporate gravity without making special assumptions in the theories.
Any theory that claims to be the correct theory of gravity must
explain these accelerating cosmologies as well as inflation.
The most plausible candidate for the consistent theory of quantum
gravity is superstrings or M-theory. So it is very important to see
how one can get a good model from these theories.

When one tries to consider this problem in string theory, one is faced
with various problems. First of all, string theories on time-dependent
backgrounds are notoriously difficult to deal with. Attempts in this
direction are made in Refs.~\refcite{CC1} and \refcite{LMS} using
the simplest orbifolds, and cosmological implications are studied.
Other approaches to strings on time-dependent backgrounds include
Refs.~\refcite{HS1,CLO}. This subject is reviewed in Ref.~\refcite{CC}.
However, it is rather difficult to discuss other time-dependent
backgrounds in the context of quantum string theories.
So it is more practical to approach the problem using the low-energy
effective supergravities.

Even within the supergravities, as we shall first summarize in the next
section, there is a no-go theorem that de Sitter solutions with
accelerating expansion are not obtained at the stationary minima in
superstring or supergravity theories. We must somehow try to evade this
theorem by relaxing some assumptions involved in the proof of
the theorem. Several attempts in this direction are well reviewed in
Refs.~\refcite{B} and \refcite{KV}, and the present paper also summarizes
various proposals, focusing mainly on time-dependent solutions to the
(generalized) Einstein equations in higher-dimensional
supergravities.\cite{TW,NO,CHNO,MO,MO1}

In the string theory, there are important extended objects called branes.
Many of the proposals to derive inflation in the string context make use
of these extended objects. In this approach, one considers that our world
exists on such branes. We shall not discuss this subject much since such
attempts to derive inflation or alternatives to inflation from the dynamics
of branes are nicely summarized in Ref.~\refcite{FQ}.

This review is organized as follows.
In the next section, we start with summarizing the no-go theorem that
claims that de Sitter solutions are impossible in supergravity theories,
and assumptions involved in the proof. This clarifies what kind of
approaches enable us to evade the no-go theorem.

In Sec.~3, we first show that relaxing one of the conditions in the proof
of the no-go theorem, we can get time-dependent solutions to vacuum Einstein
equation in higher dimensions that give accelerating cosmology.

In Sec.~4, we show that this class of solutions are actually special cases
of what are known as S-brane solutions in supergravity. We first summarize
the solutions in Sec.~4.1, discuss the relation to the vacuum solution
in Sec.~4.2, study cosmological aspects of the S-brane solutions in
Sec.~4.3 and show that one cannot in general obtain e-folding large enough
to solve the cosmological problems mentioned above. We also discuss
related solutions in type II superstrings in Sec.~4.4,
and give intuitive understanding of the basic mechanism of the
accelerating behavior of the solutions and why it is difficult to
obtain large e-folding number in these models in Sec.~4.5.
This leads us to search for other solutions with larger e-foldings.

In Sec.~5, we discuss an attempt to obtain larger e-foldings in this context,
with eternal expansion by considering hyperbolic spaces both for
external and internal spaces. This class of solutions may be useful
as a models for the present accelerating cosmology.

The generic feature of the higher dimensional gravitational theories
is that they give scalar field theories coupled to gravity in the
four-dimensional point of view.
In Sec.~6, we summarize cosmological solutions in the theories with single
scalar field with exponential potential and classify possible solutions.
The case of multi-scalars is much more difficult and we only refer to
references discussing various theories of this type.

In Sec.~7, we consider another approach that avoids the no-go theorem by
considering higher order corrections to the low-energy effective theories
of M-theory. The corrections are given as a special combination of
$R^4$ terms of Lovelock type (eight-dimensional Euler density) and specific
terms containing higher derivatives. The basic equations for the theory
are relegated to Appendix. We discuss solutions in the system in Sec.~7.1,
study their stability in Sec.~7.2 and give possible scenario that can be
obtained within these solutions in Sec.~7.3.

In Sec.~8, other approaches to accelerating universe are briefly summarized.

Finally in Sec.~9, we conclude with the summary of the results reviewed in
this paper and outlook.

\section{No-go theorem}	

The Einstein equation provide the following equations:
\bea
\mbox{Raychaudhuri eq.: }&&
\frac{\ddot a}{a} = - \frac{4\pi G}{3}(\rho+3P), \\
\mbox{Friedmann eq.: }&&
\Big(\frac{\dot a}{a}\Big)^2 + \frac{k}{a^2}=\frac{8\pi G}{3}\rho,
\ena
for the energy density $\rho$ and pressure $P$. The weak energy condition
requires that
\bea
\rho \geq 0.
\ena
In order to get inflation $\ddot a>0$, $\rho+3P$ must be negative
($w\equiv \frac{P}{\rho}<-\frac{1}{3}$),
which means anti-gravitation or gravitational repulsion.
This sounds physically nonsensical, but this is possible under some
circumstances. A famous example is the positive cosmological constant,
which gives
\bea
\rho = \Lambda, \quad
P=-\Lambda,
\ena
so that indeed we have $\rho+3P=-2\Lambda<0$, giving repulsion.
Since we have
\bea
\rho+3P=2(T_{00} -\frac12 g_{00} T^\lambda{}_\lambda)
=\frac{1}{4\pi G}R_{00},
\ena
Raychaudhuri equation reduces to
\bea
\frac{\ddot a}{a}=-\frac13 R_{00}.
\ena
So in order to have inflation, we must have $R_{00}<0$, i. e.
{\it strong energy condition must be violated.}\cite{Gibbons}

However it is not violated by eleven-dimensional supergravities
or any ten-dimensional supergravity theories corresponding to
the low-energy effective theories for M/superstring theories.
Moreover, if the higher-dimensional stress tensor satisfies the strong energy
condition, then so does the lower-dimensional stress tensor.
All of this implies that solutions of accelerated expansion are impossible
in supergravities as the low-energy effective theories of M/superstrings.
Another and more detailed argument for this no-go theorem is given below.

\subsection{Assumptions}

We consider $D(>2)$-dimensional gravity, which is compactified on
$d$ dimensions.\cite{MN} $D$-dimensional indices are denoted by $M,N,L, \ldots$
whereas $d$-dimensional ones by $\mu,\nu,\rho, \ldots$ and the rest
by $m,n,l, \ldots$.

We assume the following:
\begin{enumerate}
\item
Gravitational interactions involve no higher derivative terms.
\item
\label{ass2}
Potential is not positive definite.
\item
All the massless fields in the theory have positive kinetic terms.
\item
$d$-dimensional Newton constant is finite.
\end{enumerate}

These are the assumptions in proving the no-go theorem.
The second point may appear strange, so let us explain why this is natural.
Take the metric of our $D$-dimensional space as
\bea
ds^2 = e^{-\frac{2(D-d)}{d-2}\phi} ds_d^2 + e^{2\phi} d\Sigma_{D-d,\s}^2,
\label{reduction}
\ena
where $\s$ denotes the sign of the curvature of the internal space.
This is chosen such that the resulting $d$-dimensional theory involves
the ordinary Einstein term without multiplicative factors. This is known
as the Einstein frame in $d$ dimensions. Substituting \p{reduction} into
the Einstein action, we get the effective potential in $d$ dimensions
(see Sec.~\ref{intuitive}):
\bea
V=-\s \frac{(D-d)(D-d-1)}{2} \exp\Big[-\frac{2(D-2)}{d-2}\phi \Big].
\ena
Usually the internal space is chosen to be compact space with positive
curvature $\s=+1$, which means that the potential is negative.
The potential minimum is 0 with the field value at infinity for $\e=-1$
and for any value of field for $\e=0$.
With the presence of matter fields, the minima always take zero or negative
values, giving Minkowski or anti-de Sitter spaces. This is probably closely
related to the fact that supersymmetry is possible only in flat Minkowski
or anti-de Sitter spaces but not in de Sitter space.

\subsection{Proof}

We write $D$-dimensional Einstein equation:
\bea
R_{MN} = T_{MN} -\frac{1}{D-2} g_{MN} T^L{}_L,
\label{ein}
\ena
and the metric
\bea
ds^2 = \Omega^2(y)( dx_d^2 + \hat g_{mn} dy^m dy^n),
\ena
where the first line element $dx_d^2 = \xi_{\mu\nu} dx^\mu dx^\nu$ denotes
Minkowski or de Sitter space. From the Einstein equation~\p{ein}, we have
\bea
R_{\mu\nu} &=& R_{\mu\nu}(\xi) -\xi_{\mu\nu}(\hat \nabla^2\log \Omega
+ (D-2)(\hat \nabla\log\Omega)^2) \nn
&=& T_{\mu\nu} -\frac{1}{D-2} \Omega^2 \xi_{\mu\nu}T^L{}_L,
\label{ein1}
\ena
where the hat means that the contraction is made by $\hat g_{mn}$.
Contracted with $\xi$, Eq.~\p{ein1} gives
\bea
R(\xi) -d(\hat\nabla^2\log\Omega +(D-2)(\hat \nabla\log \Omega)^2)
= \Big(T^\mu{}_\mu -\frac{d}{D-2}T^L{}_L \Big) \Omega^2,
\ena
and hence
\bea
\hat\nabla^2\log\Omega +(D-2)(\hat \nabla\log \Omega)^2
&=& \frac{1}{d}\Big[ R(\xi) +\Omega^2 \Big(-T^\mu{}_\mu +\frac{d}{D-2}T^L{}_L
\Big)\Big],
\label{ein2}
\ena
where the energy-momentum tensors on the rhs are contracted with the
$D$-dimensional metric. If we define
\bea
\tilde T \equiv -T^\mu{}_\mu +\frac{d}{D-2}T^L{}_L,
\ena
this is non-negative. This can be verified as follows. If we have potential
$V$ for matter fields, we have $T_{MN} \sim -V g_{MN}$ and so
\bea
\tilde T = Vd -\frac{d}{D-2}DV = -\frac{2d}{D-2} V \geq 0,
\ena
where the last equality follows from our assumption~(\ref{ass2}).
If we have $n$-form fields, the energy momentum tensor takes the form
\bea
T_{MN} =F_{ML_1 \cdots L_{n-1}} F_N{}^{L_1 \cdots L_{n-1}}
-\frac{1}{2n} g_{MN} F^2,
\ena
which gives
$T^\mu{}_\mu = F_{\mu L_1 \cdots L_{n-1}} F^{\mu L_1 \cdots L_{n-1}}
-\frac{d}{2n}F^2$. Hence
\bea
\tilde T = - F_{\mu L_1 \cdots L_{n-1}}F^{\mu L_1 \cdots L_{n-1}}
+ \frac{d}{D-2}\Big(1-\frac{1}{n}\Big) F^2.
\label{trace}
\ena
Here all the indices to $F$ belong to the internal space, or if $n \geq d$
to all the $d$ dimensions and part of internal space (otherwise the isometry
of $R^d$ or $dS^d$ is broken). They separately contribute to $\tilde T$.
In the former case, $F^2 \geq 0$. It follows from \p{trace} that
$\tilde T \geq 0$ for $n>1$ and $\tilde T =0$ for $n=1$.
In the latter case, $F^2<0$ and $F_{\mu L_1 \cdots L_{n-1}}
F^{\mu L_1 \cdots L_{n-1}} =\frac{d}{n} F^2$. It again follows
\bea
\tilde T
&=& \Big[ -\frac{d}{n}+\frac{d}{D-2} \Big(1-\frac{1}{n} \Big)\Big] F^2 \nn
&=& -\frac{d(D-2-n+1)}{n(D-2)}F^2 \geq 0.
\ena
Consequently we have in general
\bea
\tilde T \geq 0.
\ena
Combined with Eq.~\p{ein2} and the assumption that our $d$-dimensional
space is Minkowski or de Sitter with non-negative scalar curvature,
this means that
\bea
\Omega^{D-2}\hat\nabla^2\Omega^{D-2} \geq 0.
\ena
The equality is true only for Minkowski space. Integrating this
over our internal space, we get
\bea
\int d^{D-d}y \sqrt{-\hat g} (\hat\nabla \Omega^{D-2})^2 \leq 0,
\ena
where we have made partial integration. The lhs is a positive-semi-definite
quantity so that this is valid only if $\Omega$ is constant and the equality
holds. This implies that the rhs of \p{ein2} vanishes and hence
de Sitter space is not allowed since the second term is positive.
Note that at this last step, we have assumed that there is no contribution
from surface terms upon partial integration. If the manifold has (singular)
boundary which produces surface contributions, the no-go theorem breaks down.

\subsection{How to avoid the theorem}
\label{avoid}

Given the confirmation of the accelerated expansion of our universe both
at early time and present,\cite{WMAP} we have to find out how this no-go
theorem may be evaded. This can be done if various assumptions, implicit
and explicit, are relaxed. Here are some possibilities.
\begin{itemize}
\item
It is assumed that the size of the internal space is time-independent.
We can try to introduce the time-dependence of the internal space.
\item
We can consider higher-derivative ``quantum correction'' terms which
are known to exist in M-theory and superstrings.
\item
We give up the compact and smooth condition on the internal space without
boundary.
\end{itemize}

The first possibility leads to the recent progress using time-dependent
solutions, and this approach is discussed in detail in this paper.
The result is that interesting solutions are obtained, but the expansion
is not enough for the inflation at early universe. However it may be useful
for the present accelerating expansion though there remain some problems to
be overcome to achieve this.

Alternatively, we can consider higher-derivative
``quantum correction'' terms which are known to exist in M-theory and
superstrings. We shall show that indeed this possibility leads to a
very promising direction of achieving inflation.

If, however, we accept supergravities without such higher order corrections,
then we must give up the compact and/or no-boundary condition
on the internal space, but this would lead to continuous spectrum in
four dimensions, which is not desirable. Compact space with boundary
is likely to suffer from nonresolvable singularities.
This possibility is not explored in this paper.

\section{Accelerating Cosmology from Vacuum Einstein Equation}

The first example of accelerating cosmology in the context of
higher-dimensional (super)gravity was obtained simply as a solution
to the vacuum Einstein equation but with hyperbolic internal space.\cite{TW}
(The use and relevance of the compact hyperbolic manifolds for the internal
space was first noted in Ref.~\refcite{KMST}. Similar time-dependent solutions
for flat internal space were discussed earlier in Ref.~\refcite{Popes}.)

It is convenient to write the $(4+n)$-dimensional solution as
\bea
ds^2 = \d^{-n}(t) ds_E^2 + \d^2(t) d\Sigma_{n,\s}^2,
\label{dsol}
\ena
where $n$ is the dimension of the internal spherical ($\s=+1$),
flat ($\s=0$) or hyperbolic ($\s=-1$) spaces, whose line elements are
$d\Sigma_{n,\s}^2$, and
\bea
ds_E^2 = -S^6(t) dt^2 + S^2(t) d{\bf x}^2,
\label{4sol}
\ena
describes the 4-dimensional spacetime. The form of \p{dsol} is chosen
such that the metrics in \p{4sol} are in the Einstein frame. The solution
is given as
\bea
\d(t) &=& e^{-3t/(n-1)} \left(\frac{\sqrt{3(n+2)/n}}{(n-1)
\sinh(\sqrt{3(n+2)/n}\; |t|)}\right)^{\frac{1}{(n-1)}}, \nn
S(t) &=& e^{-(n+2)t/2(n-1)} \left(\frac{\sqrt{3(n+2)/n}}{(n-1)
\sinh(\sqrt{3(n+2)/n}\; |t|)}\right)^{\frac{n}{2(n-1)}}.
\label{vfactor}
\ena
with hyperbolic internal space.

If we take the time coordinate $\eta$ defined by
\bea
d\eta= S^3(t) dt,
\label{time}
\ena
the metric~\p{4sol} describes a flat homogeneous isotropic universe with
scale factor $S(t)$, and $\d(t)$ gives the measure of the size of internal
space. The condition for expanding 4-dimensional universe is that
\bea
\frac{dS}{d\eta}>0.
\label{cond1}
\ena
Accelerated expansion is obtained if, in addition,
\bea
\frac{d^2S}{d\eta^2}>0.
\label{cond2}
\ena
It has been shown that these can be satisfied for $n=7$ and for certain
period of negative $t$ (with the convention $t_1=0$) which is the period
that our universe is evolving ($t<0$ and $t>0$ are two disjoint possible
universes).\cite{TW}
We note that in these solutions, accelerated expansion is obtained
only for the hyperbolic internal space; this property is lost
for flat or compact internal spaces. We shall show that this problem
is circumvented by solutions with flux, known as S-brane solutions.

\section{Accelerated Cosmologies from S-branes}

The solution in the previous section can be obtained from the
S-brane solutions in supergravities.\cite{NO}
S-branes are time-dependent solutions in supergravities, originally
considered in connection with tachyon condensation and
decay.\cite{Sbrane,Sbrane1,Sbrane2,Ohtas,Ivas,Sen1,BW}

\subsection{S-branes in supergravities}

Let us consider the $D$-dimensional gravity coupled to a
dilaton $\phi$ and $m$ different $n_A$-form field strengths:
\bea
I = \frac{1}{16 \pi G_D} \int d^D x \sqrt{-g} \left[
R - \frac{1}{2} (\pa \phi)^2 - \sum_{A=1}^m \frac{1}{2 n_A!} e^{a_A \phi}
F_{n_A}^2 \right].
\label{act}
\ena
This action describes the bosonic part of $D=11$ or $D=10$ supergravities;
we simply drop $\phi$ and put $a_A=0$ and $n_A=4$ for $D=11$, whereas we
set $a_A=-1$ for the NS-NS 3-form and $a_A=\frac{1}{2}(5-n_A)$ for forms coming
from the R-R sector.
The field strength for an electrically charged S$q$-brane is given by
\bea
F_{t \a_1 \cdots \a_{q+1}} = \e_{\a_1 \cdots \a_{q+1}} \dot E, \hs{3}
(n_A = q+2),
\label{ele}
\ena
where $\a_1, \cdots ,\a_{q+1}$ stand for the tangential directions to
the S$q$-brane. The magnetic case is given by
\bea
F^{\a_{q+2} \cdots \a_p a_1 \cdots a_{n}} = \frac{1}{\sqrt{-g}}
e^{-a\phi} \e^{t \a_{q+2} \cdots \a_p a_1 \cdots a_n} \dot E,
\hs{3} (n_A = D-q-2)
\label{mag}
\ena
where $a_1, \cdots, a_{n}$ denote the coordinates of the $n$-dimensional
hypersurface $\Sigma_{n,\s}$.

In Ref.~\refcite{Sbrane1} a single S-brane solution was given, and in
Ref.~\refcite{Ohtas} general orthogonally intersecting solutions were derived
by solving field equations. Solutions restricted to a single S$q$-brane
with $(q+1)$-dimensional world-volume in $p$-dimensional space are
(hereafter the subscript $A$ is not necessary and is dropped)
\bea
\label{oursol}
ds_D^2 &=& [\cosh\tilde c (t-t_2)]^{2 \frac{q+1}{\Delta}}
\Bigg[ e^{2c_0 t+2c_0'} \left\{ - e^{2ng(t)} dt^2
+ e^{2g(t)}d\Sigma_{n,\s}^2\right\} \nn
&& \hs{20} +\; \sum_{\a=1}^p [\cosh\tilde c(t-t_2)]^{- 2
\frac{\c^{(\a)}}{\Delta}} e^{2 \tilde c_\a t+2c_\a'} (dx^\a)^2\Bigg], \\
E &=& \sqrt{\frac{2(D-2)}{\Delta}}\frac{e^{\tilde c(t-t_2)
-\e a c_\phi'/2+\sum_{\a\in q} c_\a'}}{\cosh \tilde c(t-t_2)},\quad
\tilde c = \sum_{\a\in q} c_\a-\frac{1}{2} c_\phi \e a, \nn
\phi &=& \frac{(D-2)\e a}{\Delta} \ln \cosh\tilde c(t-t_2)
+\tilde c_\phi t + c_\phi',
\label{oursol1}
\ena
where $D=p+n+1$ and $\e= +1 (-1)$ corresponds to electric (magnetic) fields.
The coordinates $x^\a, (\a=1,\ldots, p)$ parametrize the $p$-dimensional
space, within which $(q+1)$-dimensional world-volume of S$q$-brane is
embedded, and the remaining coordinates of the $D$-dimensional spacetime
are the time $t$ and coordinates on compact $n$-dimensional spherical
($\s=+1$), flat ($\s=0$) or hyperbolic ($\s=-1$) spaces.
We have also defined
\bea
\label{res2}
&& \Delta = (q + 1) (D-q-3) + \frac{1}{2} a^2 (D-2), \nn
&& \c^{(\a)} = \left\{ \begin{array}{l}
D-2 \\
0
\end{array}
\right.
\hs{5}
{\rm for} \hs{3}
\left\{
\begin{array}{l}
x_\a \hs{3} {\rm belonging \hs{2} to} \hs{2} q{\rm -brane} \\
{\rm otherwise}
\end{array},
\right.
\ena
and
\bea
g(t) = \left\{\begin{array}{ll}
\frac{1}{n-1} \ln \frac{\b}{\cosh[(n-1)\b(t-t_1)]} & :\s=+1, \\
\pm \b(t-t_1) & :\s=0, \\
\frac{1}{n-1} \ln \frac{\b}{\sinh[(n-1)\b|t-t_1|]} & :\s=-1,
\end{array}
\right.
\label{ints}
\ena
$\b, t_1, t_2$ and $c$'s are integration constants which satisfy
\bea
&& c_0 = \frac{q+1}{\Delta}\tilde c -\frac{\sum_{\a=1}^{p} c_\a}{n-1}, \;\;
c_0' = -\frac{\sum_{\a=1}^{p} c_\a'}{n-1}, \nn
&& \tilde c_\a = c_\a - \frac{\c^{(\a)}-q-1}{\Delta}\tilde c, \;\;
\tilde c_\phi = c_\phi + \frac{(D-2)\e a}{\Delta}\tilde c.
\label{intconst1}
\ena
These must further obey the condition
\bea
\frac{1}{n-1}\left(\sum_{\a=1}^{p} c_\a\right)^2
+ \sum_{\a=1}^{p} c_\a^2 + \frac{1}{2} c_\phi^2= n(n-1) \b^2.
\label{condconst}
\ena
The free parameters in our solutions are $c_\a, c_\a' (\a=1,\cdots,p),
c_\phi, c_\phi',t_1$ and $t_2$.
The time derivative of $E$ gives the field strengths of antisymmetric
tensor and in our convention they are given as
\bea
\left. \begin{array}{c}
e^{a\phi}*\! F \\
F
\end{array}
\right\}
=\tilde c \;\sqrt{\frac{2(D-2)}{\Delta}}
e^{-\sum_{\a\in q}c'_\a+\e ac_\phi'/2} dx^{\a_{q+2}}\wedge \cdots \wedge
dx^{\a_p}\wedge \;\mbox{Vol}(\Sigma_{n,\s}),
\ena
for electric (first line) and magnetic (second line) fields, where
Vol$(\Sigma_{n,\s})$ is the unit volume form of the hypersurface
$\Sigma_{n,\s}$ and $*$ represents dual.
We can check that $\sqrt{\frac{2(D-2)}{\Delta}}=1$ for SM- and SD-branes.

For the general S2-brane obtained from the solution~\p{oursol} by putting
$p=q+1=3, c\equiv c_1=c_2=c_3, c'\equiv c_1'=c_2'=c_3'$, we find that
it takes the form~\p{dsol} and \p{4sol} with
\bea
\d(t) &=& [\cosh \tilde c(t-t_2)]^{3/\Delta} e^{g(t) + c_0 t + c_0'}, \nn
S(t) &=& [\cosh \tilde c(t-t_2)]^{(n+2)/2\Delta} e^{ng(t)/2 +(n+2)(c_0 t
+ c_0')/6},
\label{genfactor}
\ena
where
\bea
&& \tilde c = 3c-\frac{1}{2} c_\phi \e a, \quad
c_0 = \frac{3}{\Delta}\tilde c - \frac{3}{n-1}c, \quad
c_0' =-\frac{3}{n-1}c', \nn
&& \b = \sqrt{\frac{3(n+2)}{n(n-1)^2} c^2 + \frac{1}{2n(n-1)}c_\phi^2}.
\ena

\subsection{Vacuum solution}

The relation between $\tilde c$ and $c_\a$ and $c_\phi$ in Eq.~\p{oursol1}
is derived\cite{Ohtas} under the assumption that we have the independent field
strengths $F$. In the absence of these, we can disregard this relation
and set $\tilde c$ to zero.
It is then easy to see that the solution~\p{oursol} reproduces
\p{dsol}-\p{vfactor} for $p=q+1=3, \s=-1, c=1, c'=0$ without dilaton
($c_\phi=0$). The scale factor is simply~\p{genfactor} with
$\tilde c=0$ which coincides with \p{vfactor}.

The condition of the expansion~\p{cond1} for the vacuum solution~\p{vfactor}
is\cite{TW}
\bea
n_1(t) \equiv -1-\sqrt{\frac{3n}{n+2}}\coth\left(\sqrt{\frac{3(n+2)}{n}}\;
c(t-t_1)\right)>0,
\label{cond01}
\ena
where we have also included parameters $c,c'$ and $t_1$.
The condition~\p{cond2} gives
\bea
\frac{3(n-1)}{(n+2)\sinh^2[\sqrt{3(n+2)/n}\; c(t-t_1)]} - n_1^2(t) > 0.
\label{cond02}
\ena
The parameter $t_1$ and $c$ can be absorbed into the shift and rescaling
of the time $t$. Hence without loss of generality, we can set $t_1=0$ and
$c=1$ (changing $c$ gives the change in the scale of time).
There is a singularity in $S(t)$ at $t=0$, but the time $\eta$ run from 0
to infinity while $t$ runs from $-\infty$ to 0, which is an infinite
future for any event with $t<0$ and hence the evolution of our universe
can be restricted to $t<0$.

We find for $n=7$ that there is a certain period of negative time
that the conditions~\p{cond01} and \p{cond02} are satisfied.\cite{TW}
The period of the accelerated expansion can be adjusted by changing the
constant $c$, but this does not affect the resulting expansion factor.
The scale factor vanishes in the infinite past, but diverges in the infinite
future.

The expansion factor $A$ during the accelerated expansion is given by the
ratio of $S(t)$ at the starting time $T_1$ and ending time $T_2$ of the
acceleration. The expansion factor, which does not change if the constant
$c$ is changed, is found to be
\bea
A =\frac{S(T_2)}{S(T_1)} \simeq 2.91.
\ena
This value is too small to explain the cosmological problems.
Note that there is no parameter to improve $A$ here.

The behavior of the size of the internal space is also examined.
When the acceleration starts, $\d(t)$ shrinks but eventually starts expanding,
and the ratio of the sizes during the accelerated expansion is 2.18.
As observed for SM2-brane case,\cite{W} there is no stable point in the
size of the internal space, and in the infinite future and past its
size goes to infinity. This seems to be the common problem in the hyperbolic
compactification.\cite{NSS} We shall find this behavior in other cases.

Other internal spaces are also examined, but without any adjustable
parameter it is found that neither flat nor spherical spaces give accelerating
cosmologies as long as the vacuum Einstein equation is considered.\cite{NO}
Product spaces for the internal spaces are also examined, and suggestion
for large e-foldings is given.\cite{CHNW} However, there has not been given
any concrete model with large e-folding. Generalization to other dimensions
is also considered in Ref.~\refcite{Ito}.

\subsection{SM2-brane}

The SM2-brane in M-theory can be obtained from \p{oursol} by putting
$p=q+1=3, \e=+1, a=0, c_\phi=0$ without dilaton and $\Delta =3(n-1)$.
We also put $c\equiv c_1=c_2=c_3, c'\equiv c_1'=c_2'=c_3'$ and then
$\b=\frac{1}{n-1}\sqrt{\frac{3(n+2)}{n}}\; c$ is determined from \p{condconst}.
The solution~\p{oursol} then gives
\bea
ds_d^2 &=& [\cosh 3c(t-t_2)]^{2/(n-1)} \Bigg[ -e^{2ng(t)-6c'/(n-1)} dt^2
+ e^{2g(t)-6c'/(n-1)}d\Sigma_{n,\s}^2 \nn
&& \hs{20} +\; [\cosh 3c(t-t_2)]^{- 2(n+2)/3(n-1)} e^{2c'} d{\bf x}^2\Bigg].
\ena
This is the universe~\p{dsol} and \p{4sol} with
\bea
\d(t) &=& [\cosh3c(t-t_2)]^{1/(n-1)} e^{g(t) -3c'/(n-1)}, \nn
S(t) &=& [\cosh3c(t-t_2)]^{(n+2)/6(n-1)} e^{ng(t)/2 -(n+2)c'/2(n-1)}.
\label{mfactor}
\ena
We now discuss three internal spaces~\p{ints} separately.

\subsubsection{Hyperbolic internal space}

The conditions~\p{cond1} and \p{cond2} for $n=7$ of our interest and
hyperbolic internal space $\s=-1$ are (again shifting the time to set $t_1=0$)
\bea
\label{cond11}
&& n_2(t) \equiv \frac{3}{4} \tanh[3c(t-t_2)]-\frac{\sqrt{21}}{4}
\coth(3\sqrt{3/7}ct)>0,\\
&& \frac{9}{8}\left(\frac{1}{\cosh^2[3c(t-t_2)]}
+ \frac{1}{\sinh^2(3\sqrt{3/7}ct)}\right) - n_2^2(t) > 0.
\label{cond12}
\ena

Here we can again consider that our universe evolves only for $t<0$.
The qualitative behaviors of the left hand side of these Eqs. for $c=1$
and $t_2=0$ are similar to those in the previous S-brane case, and
we again find that there is a certain period of
negative time that these conditions are satisfied.\cite{NO}

The expansion factor is found to be\cite{W,NO}
\bea
A \simeq 2.17,
\ena
which is too small to explain the cosmological problems.
However, here is a parameter $t_2$ in contrast to the vacuum solution,
whose effect is examined. It is found that
the typical behavior for positive $t_2$ is basically the same as $t_2=0$
case, but the period of the accelerated expansion changes slightly.
For example, for $t_2=1$, we find
the value of the expansion factor during the accelerated expansion
improves:
\bea
A \simeq 3.13.
\ena
This is still not enough improvement for cosmological applications.
Increasing the value of $t_2$ does not affect the numerical value of $A$
much. The behavior for negative $t_2$ is also basically the same, but the
expansion factor is worse; one typically gets $A \sim 1.4$.

The behavior of the size of the internal space for $t_2=0$ is also similar.
As observed in Ref.~\refcite{W} there is no stable point in the size of the
internal space, and in the infinite future its size goes to infinity.
There is no significant change in this behavior if we change the parameter
$t_2$.

\subsubsection{Flat internal space}

The conditions~\p{cond1} and \p{cond2} for $n=7$ and flat internal
space $\s=0$ are
\bea
\label{cond21}
&& n_3(t) \equiv \frac{3}{4} \tanh[3c(t-t_2)]+\frac{\sqrt{21}}{4}>0, \\
&& \frac{9}{8}\frac{1}{\cosh^2[3c(t-t_2)]} - n_3^2(t) > 0,
\label{cond22}
\ena
where we have chosen the plus sign in Eq.~\p{ints} since minus sign cannot
give expanding universe.

Here since $S(t)$ does not have any singularity and is positive, the time
$t$ runs from $-\infty$ to $+\infty$ while the time $\eta$ runs from 0
to $+\infty$ monotonously. The left hand side of these Eqs. for $c=1$ and
$t_2=0$ are examined and we again find that there is a certain period of
negative time that these conditions are satisfied.\cite{NO}

The expansion factor in this case is found to be
\bea
A \simeq 1.35,
\ena
which is again too small to explain the cosmological problems.
The conditions~\p{cond21} and \p{cond22} depend only on $t-t_2$,
so changing $t_2$ simply shifts the evolution of the spacetime and
does not give any difference.

The behavior of the size of the internal space is also similar.
There is no stable point in the size of the internal space, and in the
infinite future its size goes to infinity.

\subsubsection{Spherical internal space}

The conditions~\p{cond1} and \p{cond2} for $n=7$ and spherical internal
space $\s=+1$ are (again shifting the time to set $t_1=0$)
\bea
\label{cond31}
&& n_4(t) \equiv \frac{3}{4} \tanh[3c(t-t_2)]
- \frac{\sqrt{21}}{4} \tanh(3\sqrt{3/7}\; ct)>0, \\
&& \frac{9}{8}\left(\frac{1}{\cosh^2[3c(t-t_2)]}
- \frac{1}{\cosh^2(3\sqrt{3/7}\; ct)}\right) - n_4^2(t) > 0.
\label{cond32}
\ena
The time ranges of $\eta$ and $t$ are the same as the flat internal space.
It was first thought that there is no period of accelerated expansion.
However when the parameter $t_2$ is changed, the results change.
For $c=1$ and $t_2=-1$, we find that there is a certain period of
negative time that the conditions~\p{cond31} and \p{cond32} are satisfied,
though the universe begin contraction after some positive time.
The scale factor $S(t)$ in~\p{mfactor} contracts on both ends
$t\to \pm \infty$.
The expansion factor is
\bea
A \simeq 1.35,
\ena
which is again too small to explain the cosmological problems.
We have also checked that the typical behavior for negative $t_2$ is again
basically the same as $t_2=-1$ case, but the period of the accelerated
expansion and the value of the expansion factor during the accelerated
expansion change.

A different behavior is observed for positive $t_2$. We find the
acceleration occurs while the universe is already contracting for $t>0$.
There is no stable point in the size of the internal space, and in the
infinite future its size goes to infinity. There is no significant change
in this behavior if we further change the parameter $t_2$.

The value of the expansion factor during the accelerated expansion in all
these solutions are so small that they cannot give inflationary models
at the early epoch. How about the present accelerating cosmology?
The size of the internal space of these models is examined in
Ref.~\refcite{GKL} and it was concluded that it is too large to give
a four-dimensional model of the present accelerating universe.
Thus we would need some mechanism to keep the size of the internal space
small, or consider a kind of warp factors to render the large extra dimensions
harmless.\cite{RS} It is desirable to find modification of these solutions
without invoking some special objects, which need further explanation why
it is natural to consider such objects. Such approach would be similar to
introducing scalar field with suitable potential, and in a sense contradicts
our spirit to use our fundamental theory. We hope to derive an accelerating
cosmologies in a well-defined setting without the need of further
explanation, and this problem of stabilization of the size of the internal
space is a subject left for future study.

\subsection{SD2-brane}

In this section, we discuss accelerating cosmologies from
SD2-brane.\cite{NO,roy}

We first show how the SD2-brane can be obtained by dimensional reduction
from SM2-branes.\cite{NO} Though it is possible to examine general
dimensions, we restrict ourselves to $D=11$ here.

The ten-dimensional metric in the string frame is obtained from the
eleven-dimensional one by the relation\cite{Ohtam}
\bea
ds_{11}^2 = e^{-2\phi/3} ds_{10,s}^2 + e^{4\phi/3} dx_{10}^2.
\label{11metric}
\ena
In our solutions~\p{oursol}, we consider SM2-brane by taking $q=2$,
but we set $p=4$ with one extra coordinate outside SM2-brane, in which
direction we make dimensional reduction. We choose $x_1,x_2,x_3$ to be
the SD2-brane world-volume and $x_4$ the direction of dimensional reduction.
We further set $c\equiv c_1=c_2=c_3, c' \equiv c_1'=c_2'=c_3'$ but leave
$c_4$ and $c_4'$ arbitrary. There is no dilaton which comes from the metric
as~\p{11metric} upon dimensional reduction, and we put $a=c_\phi=0$.
This gives
\bea
\phi &=& \frac{1}{4} \ln [\cosh 3c (t-t_2)] + \frac{3}{4} (c+2c_4)t
+ \frac{3}{2} c'_4, \nn
ds_{10,s}^2 &=& e^{2\phi/3} [\cosh 3c (t-t_2)]^{1/3}\Bigg[
e^{-(c+2c_4)t/5 - 2(3c'+c_4')/5} \left\{ - e^{12 g(t)} dt^2
+ e^{2g(t)}d\Sigma_{6,\s}^2\right\} \nn
&& \hs{20} +\; [\cosh 3c(t-t_2)]^{-1} e^{2c'} d{\bf x}^2 \Bigg].
\ena
In the Einstein frame $ds_{10,E}^2 = e^{-\phi/2} ds_{10,s}^2$, this reduces to
\bea
ds_{10,E}^2 &=& [\cosh 3c (t-t_2)]^{3/8}\Bigg[ e^{-3(c+2c_4)t/40-3(8c'
+c_4')/20} \left\{ - e^{12 g(t)} dt^2 + e^{2g(t)}d\Sigma_{6,\s}^2
\right\} \nn
&& \hs{20} +\; [\cosh 3c(t-t_2)]^{-1} e^{(c+2c_4)t/8+(8c'+c_4')/4} d{\bf x}^2
\Bigg].
\ena

The scale factor $S(t)$ and $\d(t)$ are given as
\bea
\label{dfactor}
\d(t) &=& [\cosh (16\a/5)(t-t_2)]^{3/16} e^{g(t) -3c_1t/64 - 3\tilde c'/4},\nn
S(t) &=& [\cosh (16\a/5)(t-t_2)]^{1/4} e^{3g(t) -c_1 t/16 - \tilde c'},\\
\phi &=& \frac{1}{4}\ln[\cosh(16\a/5)(t-t_2)] +\frac{15}{16}c_1 t +c_\phi',
\nonumber
\ena
where we have also put
\bea
\tilde c'= \frac{8c'+c_4'}{10}, \quad
c_\phi' = \frac{3}{2} c_4'.
\ena
On the other hand, if we consider the SD2-brane with $D=10, q=2, p=3, n=6,
a=\frac{1}{2},\e=+1$ in Eq.~\p{oursol}, we get precisely \p{dfactor}
with different parametrization.
Continuing this reduction, one should leave the constants $c_\a$ arbitrary
in order to get general lower-dimensional solutions.

The scale factors~\p{dfactor} almost coincide with those for SM2-brane
in~\p{mfactor} if we choose $n=6$, though the power of the $\cosh$ is
slightly different and there is an additional $c_1 t$ term in the
exponent. However, the difference is small and one may expect
that qualitative features remain the same.
In fact this is explicitly confirmed in Ref.~\refcite{NO}.

One potential subtlety in this picture of SD2-brane is that
the string coupling is given by the dilaton as $e^\phi$,
and it may diverge for the solution~\p{dfactor} for certain range of
parameters. If this happens, quantum effects dominate and we cannot simply
trust the result. This can be avoided for hyperbolic internal space
but it becomes more serious problem for other internal space.
For instance, for flat internal space it is found that
the dilaton~\p{dfactor} always gives strong string coupling somewhere for
large $|t|$ whatever the choice of the parameters are, and it is not clear
if this classical analysis is valid.
The strong coupling limit is the one in which eleventh dimension
becomes large.\cite{WI} This case is better described in the SM2-brane
picture, with large eleventh dimension. This is what we have already analyzed
in the preceding section.

\subsection{Intuitive approach using effective four-dimensional potential}
\label{intuitive}

So far we have seen that the accelerating behavior is possible for
time-dependent internal space, and this is possible only for the hyperbolic
internal space for vacuum solution, but it is found that S-brane
solutions give this behavior for flat and compact spaces. Why is this so?
In fact, there is a very intuitive way to understand this, using
the effective four-dimensional potential.\cite{EG}

Let us consider the product space of $d$-dimensional universe and
$\mathbb M_1\times\mathbb M_2\times\mathbb M_3$ with dimensions $m_1, m_2,
m_3$ and signatures of curvature $\s_1, \s_2, \s_3$. Products of one or
two spaces for extra dimensions can be obtained by setting $m$'s to zero.
The ansatz of the metric for the full spacetime is
\begin{eqnarray} \label{metfors}
ds^2 = {\rm e}^{2\phi(x)} ds_{d}^2 + \sum_{i=1}^{3} {\rm e}^{2\phi_i(x)}
d\Sigma^2_{m_i,\s_i},
\end{eqnarray}
where we have chosen the Einstein frame by setting
\begin{eqnarray}
\phi= -\sum_i m_i \phi_i/(d-2).
\end{eqnarray}
The kinetic terms for the scalars corresponding to the radii of
each internal space in the effective theory are given by
\begin{eqnarray}
&& \hs{-10}
K = \frac{\rho+p}{2} = \sum_{i=1}^3
\frac{m_i(m_i+d-2)}{2(d-2)a^2} {\dot\phi_i}^2 + \sum_{i>j=1}^{3}
\frac{m_i m_j}{(d-2)a^2} \dot\phi_i \dot\phi_j - \s_0
\frac{d-2}{2a^2}.
\label{effK}
\ena
They can be diagonalized and normalized by a field redefinition
\begin{eqnarray}
\psi_1 &=& \sqrt{\frac{m_1(m_1+d-2)}{d-2}} \Big[ \phi_1 +
\frac{1}{m_1+d-2} (m_2 \phi_2 + m_3 \phi_3) \Big], \nn
\psi_2 &=& \sqrt{\frac{m_2(m_1+m_2+ d-2)}{m_1+d-2}} \Big[ \phi_2
+ \frac{m_3}{m_1+m_2+d-2} \phi_3 \Big], \\
\psi_3 &=& \sqrt{\frac{m_3(m_1+m_2+m_3+ d-2)}{m_1+m_2+d-2}}\phi_3,\nonumber
\end{eqnarray}
with the result
\begin{eqnarray}
K &=& \frac{1}{2} \sum_{i=1}^3 \dot{\psi}_i^2 - \e_0
\frac{d-2}{2a^2},
\\
V &=& - \sum_{i=1}^3 \s_i \frac{m_i(m_i-1)}{2} e^{\sum_a
2M_{ia}\psi_a} - \s_0 \frac{(d-2)^2}{2a^2},
\label{effV}
\end{eqnarray}
where the matrix $M_{ia}$ is given by
\begin{eqnarray}
\hs{-3}
M_{ia} =\! \left(\begin{array}{ccc}
-\sqrt{\frac{m_1+d-2}{(d-2)m_1}} & 0 & 0 \\
-\sqrt{\frac{m_1}{(d-2)(m_1+d-2)}} &
-\sqrt{\frac{m_1+m_2+d-2}{m_2(m_1+d-2)}} & 0 \\
-\sqrt{\frac{m_1}{(d-2)(m_1+d-2)}} &
-\sqrt{\frac{m_2}{(m_1+d-2)(m_1+m_2+d-2)}} &
-\sqrt{\frac{m_1+m_2+m_3+d-2}{m_3(m_1+m_2+d-2)}}
\end{array} \right).
\label{matrix}
\end{eqnarray}

To study the properties of the potential, it is more convenient
to define new independent fields as
\begin{eqnarray}
\vp_1 &\equiv& 2\sqrt{\frac{m_1+d-2}{m_1(d-2)}}\psi_1, \nn
\vp_2 &\equiv& 2\sqrt{\frac{m_1}{(d-2) (m_1+d-2)}} \psi_1+
2\sqrt{\frac{m_1+m_2+d-2}{m_2(m_1+d-2)}}\psi_2, \\
\vp_3 &\equiv& 2\sqrt{\frac{m_1} {(d-2)(m_1+d-2)}}\psi_1 +
2\sqrt{\frac{m_2}{(m_1+d-2)(m_1+m_2+d-2)}}\psi_2 \nn
&& + 2\sqrt{\frac{m_1+m_2+m_3+d-2}{m_3(m_1+m_2+d-2)}}\psi_3.\nonumber
\end{eqnarray}
and the effective potential (\ref{effV}) becomes\cite{CHNO}
\begin{eqnarray}
V = - \sum_{i=1}^3 \s_i \frac{m_i(m_i-1)}{2} {\rm e}^{-\vp_i} -
\s_0 \frac{(d-2)^2}{2a^2}.
\end{eqnarray}
Clearly the potential is unbounded if any one of the $\s_i$'s is
positive in that direction and positive with minimum at infinity if
the $\s_i$ is negative. The potentials for positive and negative $\s_i$
are shown by solid lines in Fig.~\ref{pot}. However, if we add contributions
from antisymmetric tensors, this is modified. For instance, the contribution
of the 4-form field in eleven-dimensional supergravity is
\begin{eqnarray}
\Delta V = b^2 \exp\Big[ -\frac{(d-1)(m_1 \vp_1 + m_2 \vp_2+m_3
\vp_3)} {m_1+m_2+m_3+d-2}\Big],
\label{flux}
\end{eqnarray}
where $b$ is a constant. Now the potential is bounded from below
if $\frac{(d-1)m_i}{m_1+m_2+m_3+d-2}>1$ for the direction $\s_i=+1$,
as shown by the dashed line in Fig.~\ref{pot}.
(There is no requirement and no minimum in the direction $\s_i=-1,0$.)
For $d=4$, this is easily satisfied. In this case, there will be local minimum
in the direction with $\s_i=+1$, but the potential minimum is always negative.

Single internal space is discussed in Ref.~\refcite{EG}, with only
one scalar field $\vp_1$, $d=4$ and $m_1=n, m_2=m_3=0$.
\begin{figure}[h]
\begin{center}
\setlength{\unitlength}{.7mm}
\begin{picture}(50,40)
\includegraphics[width=5cm]{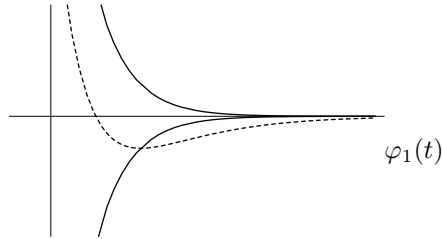}
\put(0,15){$\vp_1(t)$}
\end{picture}
\caption{Potential for scalar field for $\s=\pm 1$ with $b=0$ (solid)
and $b\neq 0,\s=+1$ (dashed).}
\label{pot}
\end{center}
\vs{-3}
\end{figure}
Suppose that the
field starts at a large value of $\vp_1>0$, with large negative velocity
$\dot\vp_1$. This is a phase of decelerated expansion. If the kinetic
energy of $\vp_1$ is large enough, the field will run up the exponential
potential hill at later time, and it will stop. Around this, the
potential dominates and the universe is in the accelerating phase.
Soon after this, the field starts rolling back down the hill and the universe
gets back to the decelerating phase.

It is clear that this behavior is obtained for the vacuum without flux
($b=0$) only if $\e=-1$, namely for hyperbolic space, whereas it
can be obtained for the S-brane solutions with flux for all kinds of
internal spaces. This is the reason why this solution is obtained
only for the hyperbolic internal space in Townsend-Wohlfarth solution.
However, it is difficult to get big expansion even in the present case
since the exponential potential is too steep to keep the field at the top
for long time. This remains true if we take the internal space to be
the product space.\cite{CHNO}
In order to have big expansion, we should have local minimum at positive
value, where our universe stays for a while and expands, and then decays to
lower value. However the no-go theorem claims that there is no such minimum.

It is also interesting that a potential with local minimum (though at
negative value) is obtained in the Einstein gravity coupled to
gauge fields, providing a mechanism for stabilizing the size of
extra dimensions. How to stabilize the size of extra dimensions
is an issue no less important than obtaining inflation. This is a
direction worthy of further exploration.

Another interesting research direction is to try to obtain inflation
in the present picture by introducing matter fields into the solutions.
This could be done, for example, by considering D-branes in the solutions.
Such solutions have recently been discussed in Ref.~\refcite{EM}.
Related models using D-branes are proposed\cite{DHHK} and have been
extensively studied. Brane and anti-brane systems are also a focus of
extensive study.\cite{BMNQ} Such systems naturally contain tachyons
which may drive accelerating behavior.\cite{tachyon}
Unfortunately it is pointed out that it is rather difficult to obtain
inflation and it can occur only at super-Planckian densities where the
effective four-dimensional field theory is not applicable.\cite{tach1}

A scalar field with a positive potential which yields an accelerating
universe has been named ``quintessence''.\cite{quint1,quint2}
Supergravity models which exhibit such properties are discussed in
Ref.~\refcite{T}, and rather systematic study of theories with a scalar
with exponential potential and connection with gauged supergravities
are presented in Ref.~\refcite{Bexp}. Possible problems with
this approach in the attempt at making sense of quantum theory of string
theories are pointed out in Ref.~\refcite{Hellerman}. More exotic models with
scalar fields with higher order kinetic terms (called ``$k$-essence'')
are shown to lead to power-law expansion,\cite{Chimento} but the problem
with this approach is how to justify the existence of such a scalar field.

\section{Eternal Expansion}

Considering the difficulty in obtaining large e-foldings in the simple S-brane
solutions, it is desirable to investigate whatever modifications to improve
it. Here we present one interesting solution which exhibits eternal
expansion.\cite{CHNO}

\subsection{An exact critical solution}

Let the higher-dimensional geometry be
$\mathbb R\times\mathbb H_d\times\mathbb H_m$.
The metric ansatz is
\begin{equation}
ds^2 = {\rm e}^{-\frac{2m}{d-2}\,\phi(t)} \left( - dt^2 + a(t)^2
\, ds_{Hd}^2 \right) + {\rm e}^{2\phi(t)} ds_{Hm}^2 \,.
\end{equation}
The size of $\mathbb H_m$ corresponds to a scalar field in the effective
theory on $\mathbb R\times\mathbb H_d$. Here we write the (non-trivial)
components of the Einstein tensor for arbitrary $\s_0$, $\s_1$:
\begin{eqnarray}
G_{00} &=& - \left[ \frac{\lambda}{2} \, \dot{\phi}^2 -
\s_1 \frac{m(m-1)}{2} \, {\rm e}^{-\frac{2\lambda}{m}
\phi(t)} - \frac{(d-1)(d-2)}{2} \left( H^2 + \frac{\s_0}{a^2}
\right) \right] \,, \label{G00}
\\
G_{xx} &=& - a^2 \left[ \frac{\lambda}{2}\, \dot{\phi}^2 +
\s_1 \frac{m(m-1)}{2} \, {\rm e}^{-\frac{2\lambda}{m}
\phi(t)} + \frac{(d-2)(d-3)}{2} \left( H^2 +
\frac{\s_0}{a^2} \right) + (d-2) \frac{\ddot{a}}{a}
\right], \label{Gxx}
\nn
G_{ii} &=& \Bigg[ \frac{\lambda}{m} \left( \ddot{\phi} + (d-1)\,H
\dot{\phi} \right) - \frac{\lambda}{2} \dot{\phi}^2 - \s_1
\frac{(m-1)(m-2)}{2} \, {\rm e}^{-\frac{2\lambda}{m}\phi(t)}
\nn
&{}&\qquad\qquad\qquad - \frac{(d-1)(d-2)}{2} \left( H^2 +
\frac{\s_0}{a^2} \right) - (d-1)\,\frac{\ddot{a}}{a} \Bigg] {\rm
e}^{\frac{2\lambda}{m}\phi(t)}\,, \label{Gii}
\end{eqnarray}
where $\lambda\equiv \frac{m(m+d-2)}{(d-2)}$. The metric on
$\mathbb R\times\mathbb H_{d-1}\times\mathbb H_m$ space takes the values
$\s_0=\s_1=-1$. Putting $d=4$ and by a change of variable
\begin{equation}
\phi = \sqrt{\frac{2}{m(m+2)}} \psi + \frac{1}{m+2}\ln(m(m-1)),
\end{equation}
we write the Friedman equation and the wave equation for $\psi$ as
\begin{eqnarray}
&3H^2 = \frac{1}{2} \dot{\psi}^2 + \frac{1}{2} {\rm e}^{-c\psi} +
\frac{3}{a^2}, \qquad c = \sqrt{\frac{2(m+2)}{m}}, \label{eqH}
\\
&\ddot{\psi} + 3 H \dot{\psi} - \frac{c}{2} {\rm e}^{-c\psi} = 0.
\label{eqpsi}
\end{eqnarray}
It is straightforward to obtain the critical (in the sense that the power
exponent of the scale factor is on the boundary to be inflationary)
solution with $\ddot{a} = 0$:
\begin{equation}\label{exactsol}
a = \sqrt{\frac{m+2}{2}} t, \qquad \psi = \frac{2}{c}\ln(t) +
\frac{1}{c}\ln\left( \frac{c^2}{8} \right).
\end{equation}

\subsection{Scalar perturbations}
\label{perturb}

A tiny perturbation to the exact solution is found to lead to
interesting behavior of an accelerating phase. Consider a small
perturbation around the
solution~(\ref{exactsol}). Let
\begin{equation}
a = a_0 + a_1, \qquad \psi = \psi_0 + \psi_1,
\end{equation}
where $a_0$ and $\psi_0$ are given by (\ref{exactsol}). It follows
that the Hubble parameter is
\begin{equation}
H = H_0 + H_1, \qquad H_0 = \frac{1}{t}, \qquad H_1 =
\frac{\dot{a}_1}{a_0} - H_0 \frac{a_1}{a_0},
\end{equation}
to the first order approximation. Expanding Eqs.~\p{eqH} and \p{eqpsi}
and keeping first order terms only,
we get
\begin{eqnarray}
&6H_0 H_1 = \dot{\psi}_0 \dot{\psi}_1 - \frac{c}{2} {\rm
e}^{-c\psi_0} \psi_1 - 6 \frac{a_1}{a_0^3},
\\
&\ddot{\psi}_1 + 3 H_0 \dot{\psi}_1 + 3 H_1 \dot{\psi_0} +
\frac{c^2}{2} {\rm e}^{-c\psi_0} \psi_1 = 0 \,,
\end{eqnarray}
along with
\begin{equation}
2 \dot{\psi_0} \dot{\psi_1} + \frac{3\ddot{a_1}}{a_0} +
\frac{c}{2} \, {\rm e}^{-c\psi_0} \psi_1 = 0 \,.
\end{equation}
These linear equations are solved with the following solutions
\begin{equation}
\label{a1phi1} a_1 = A t^n, \qquad \psi_1 = \gamma A t^{n-1},
\end{equation}
where
\begin{equation}
\gamma = \frac{3(1-n)}{2\sqrt{m}} \,; \qquad n = \pm
\sqrt{\frac{m-6}{m+2}}\,.
\end{equation}
These give real solutions only if
\begin{equation}
m > 6\,.
\end{equation}
Note that $m=6$ or $n=0$ is excluded because it is just a zero mode
corresponding to time-shift symmetry. There are solutions with eternally
accelerating expansion when $m\geq 7$, although $m=7$ is the
most interesting case since $m=7$ (together with our
spacetime 4 dimensions) is precisely the dimension in which
M-theory lives. This coincidence suggests that the approach is worth
serious consideration.

For the case $m=7$, we have
\begin{equation}
n = \frac{1}{3}, \;\; -\frac{1}{3}; \qquad \gamma =
\frac{1}{\sqrt{7}}\,,\, \frac{2}{\sqrt{7}}\,.
\end{equation}
Examining the $n=1/3$ solution we find that
\begin{equation}
a_1 = A t^{1/3}, \qquad \ddot{a} = -\frac{2A}{9 t^{5/3}},
\end{equation}
so this gives positive acceleration for $A <0$. However, as time
increases, $a_1$ grows and perturbative expansion is no longer valid.
We cannot claim eternally accelerating expansion for this
solution without further analysis. The other solution $n=-1/3$,
\begin{equation}
a_1 = A t^{-1/3}, \qquad \ddot{a} = \frac{4A}{9 t^{7/3}},
\end{equation}
gives a positive acceleration for $A > 0$. As time increases,
$a_1$ approaches to zero and our perturbative calculation remains
valid. We find eternally accelerating expansion for this case.
This behavior is also confirmed by numerical evaluation without recourse
to perturbation.

For product space compactifications, it is generally difficult to
find exact solutions for the coupled Einstein equations unless the
internal space is a product of flat spaces and at most one
nontrivial curved space or they all are of the same type.  It
would be interesting to find the exact solution corresponding to
the solution we have obtained here with eternally accelerating
expansion and see if the inflation is further increased at early
times and not just at late times.

\section{Cosmologies with exponential potentials}

In Sec.~4.4, we have seen that the cosmological solutions
can be analyzed by using the effective theories in four dimensions,
which typically contains exponential potentials of scalar fields.
This may be regarded as an alternative approach using the effective
four-dimensional theories.
In this section, we briefly discuss this subject.

The cosmological solutions of the system with a single scalar field
can be systematically studied and classified.\cite{HAL,Texp,Vexp}
The field equations to be studied are Eqs.~\p{eqH} and \p{eqpsi}.
We define a new time parameter
\bea
d\tau = e^{-\frac{c}{2}\psi} dt,
\label{newtime}
\ena
and set $a(t)=e^{\a(\tau)}$. Denoting the derivative with respect to $\tau$
by a dot, Eqs~\p{eqH} and \p{eqpsi} for hyperbolic internal space reduce to
\begin{eqnarray}
&3\dot \a^2 - \frac{1}{2} \dot{\psi}^2 = \frac{1}{2}
- \s_0 \frac{3}{a^2}e^{c\psi},\qquad
c = \sqrt{\frac{2(m+2)}{m}}, \label{eH}
\\
&\ddot{\psi} - \frac{c}{2} \dot{\psi}^2 +3 \dot \a \dot{\psi} = \frac{c}{2}.
\label{psi}
\end{eqnarray}
Note that the constant is restricted to the range
\bea
\sqrt{\frac{18}{7}} \geq c >\sqrt{2},
\label{crange}
\ena
for $m \geq 7$.
When the 4-form flux is included in M-theory, we find from Eq.~\p{flux}
that there appears a potential of the form
\bea
b^2 e^{-\frac{6}{c} \psi}.
\ena
The range of the coefficient is
\bea
\frac{6}{\sqrt{2}} > \frac{6}{c} \geq \sqrt{14},
\ena
for $m \geq 7$.

For $\s_0=0$, Eq.~\p{eH} defines a hyperbola which separates the $\s_0=-1$
and $\s_0=+1$ trajectories. We are interested in the $\dot \a>0$ branch
where we have expanding universe. Let us parametrize this branch as
\bea
\dot \a =\frac{1}{\sqrt{24}}(\xi + \xi^{-1}), \qquad
\dot \psi = \frac{1}{2}(\xi - \xi^{-1}), \qquad
(\xi>0).
\label{para}
\ena
Equation~\p{psi} then becomes
\bea
\dot\xi = \frac{1}{4}[ c + \sqrt{6} + (c - \sqrt{6}) \xi^2].
\label{ana}
\ena
We are now going to study several cases separately.

\subsection{$c<\sqrt{6}$}

In this case (which is within the range~\p{crange}), there is a fixed
point solution
\bea
\xi= \xi_0 \equiv \left(\frac{\sqrt{6}+c}{\sqrt{6}-c}\right)^\frac{1}{2}.
\ena
Since $\xi$ is a constant, Eqs.~\p{para} shows that $\a$ and $\psi$ are
linear in $\tau$:
\bea
\a =\frac{\tau}{\sqrt{6-c^2}} + \mbox{const.}, \qquad
\psi =\frac{c}{\sqrt{6-c^2}}\tau + \mbox{const.}.
\ena
We deduce from Eq.~\p{newtime} that
\bea
t \sim e^{\frac{c^2}{2\sqrt{6-c^2}}\tau},
\ena
and hence
\bea
e^{\psi} \sim t^{2/c}, \qquad
a \sim t^{2/c^2}.
\label{att}
\ena
This is an expanding solution by power law.

There are two solutions to Eq.~\p{ana}:
\bea
(i)\; \xi=\xi_0 \tanh \c \tau, \qquad
(ii)\; \xi=\xi_0 \coth \c \tau, \qquad
\c \equiv \frac{\sqrt{6-c^2}}{4}.
\ena
Only the solution $(i)$ includes $\xi=1$, hence $\dot\psi=0$. This solution
undergoes a period of accelerating expansion. We find
\bea
a^{\sqrt{6}} \propto (\cosh \c \tau)^{\lambda_-}(\sinh \c \tau)^{\lambda_+},
\qquad
e^\psi \propto (\cosh \c \tau)^{\lambda_-}(\sinh \c \tau)^{-\lambda_+},
\ena
where
\bea
\lambda_\pm = \frac{2}{\sqrt{6}\pm c}.
\ena
This solution exhibits a decelerating epoch and then passes through
a period of acceleration and finally approaches the attractor~\p{att}.
In case $(ii)$, one has
\bea
a^{\sqrt{6}} \propto (\cosh \c \tau)^{\lambda_+}(\sinh \c \tau)^{\lambda_-},
\qquad
e^\psi \propto (\cosh \c \tau)^{-\lambda_+}(\sinh \c \tau)^{\lambda_-}.
\ena

If $c=\sqrt{2}$ is allowed, the power-law solution~\p{att} gives
$a\sim t$ with zero acceleration. In this case, the above solution
approaching this asymptotically must be eternally accelerating
solution, which is similar to that found in Sec.~5.

\subsection{$c>\sqrt{6}$}

The case for $c>\sqrt{6}$ is similar. The solution to Eq.~\p{ana} is
\bea
\xi = \left(\frac{c+\sqrt{6}}{c-\sqrt{6}}\right)^{\frac{1}{2}}
\tan \omega\tau, \qquad
\omega \equiv \frac{\sqrt{c^2-6}}{4},
\ena
where $0<\tau < \pi/2$. Eqs.~\p{para} can be integrated to give
\bea
a^{\sqrt{6}} \propto (\cos\omega\tau)^{\lambda_-}(\sin\omega\tau)^{\lambda_+},
\qquad
e^\psi \propto (\cos\omega\tau)^{\lambda_-}(\sin\omega\tau)^{-\lambda_+}.
\ena
The asymptotic behavior for $\log t \to \pm \infty$ is
\bea
a \sim t^{1/3}, \qquad
e^{\psi} \sim t^{\pm 2/\sqrt{6}}.
\ena
There is also a period of acceleration.

\subsection{$c=\sqrt{6}$}

We have simply $\xi=\frac{\sqrt{6}}{2} \tau$ and
\bea
a^{3} \sim \tau^{1/2} e^{\frac{3}{8}\tau^2}, \qquad
e^{\frac{\sqrt{6}}{2}\psi} \sim \tau^{-1/2} e^{\frac{3}{8}\tau^2}.
\ena
The asymptotic behavior is power-law with logarithmic corrections.

Solutions with flux in this approach have been discussed in Ref.~\refcite{Nexp}
and general discussions are given in Ref.~\refcite{JMS}.
A more systematic method to solve field equations by using new variable
is discussed in Ref.~\refcite{Russo}.
Power-law solutions of inflationary nature with exponential potential
is discussed in Ref.~\refcite{exp2}.

\subsection{Case of multi-scalars}

The cosmological solutions with multi-scalars and hence with multi-exponential
potentials are more involved. Some interesting solutions for this case are
obtained in Refs.~\refcite{Cexp1,Cexp2,Guo,IMexp}. Solutions to exponential
potentials with pressureless baryonic matter are discussed in
Ref.~\refcite{KKexp}.
General discussions of attractors and repeller solutions are given in
Refs.~\refcite{CNR}, to which we refer the reader for more details.
{}From the string theory point of view, it is natural to consider
antisymmetric tensors together with scalar fields with exponential potential.
Solutions for such systems are studied and discussed in Ref.~\refcite{BNQT}.
Similar system of antisymmetric tensors and scalars is investigated in
Ref.~\refcite{APD} for generating large four dimensions.

\section{Generalized de Sitter Solutions from Theories with Higher-order Terms}

We note that the scale when the acceleration occurs in the time-dependent
solutions related to S-branes is basically governed by the Planck scale in
the higher (ten or eleven) dimensions. When dealing with phenomena at such
high energy, it is expected that we cannot ignore higher order corrections
to the lowest Einstein-Hilbert (EH) term in the theories at least in the
early universe. In fact there are terms of higher orders in the curvature
to the lowest effective supergravity action coming from superstrings or
M-theory. In four dimensions, many studies have been done with such
correction terms.\cite{staro}

The cosmological models in higher dimensions were also studied intensively
in the 80's by many authors.\cite{KK,WH,Maeda1,ISHI,Maeda2,MMS,Maeda3}
Higher order terms in scalar curvature were considered in
Refs.~\refcite{WH,Maeda3,EKO}. In particular, it was shown that these theories
can be mapped to a gravity system coupled with a scalar field by
a conformal transformation,\cite{WH,Maeda3} and hence are very easy to
deal with.

The higher-order curvature terms called Lovelock gravity\cite{LOV} were
also considered in higher-dimensional cosmology.\cite{DF,M,KKPV} The lowest
higher order corrections are expected to appear in this form called
Gauss-Bonnet (GB) terms which are the special combination without
ghosts.\cite{Zw} In fact, it is known that this combination appears
in the low-energy effective theories of heterotic strings,\cite{Be,hetero}
and extensive study of these theories was carried out.\cite{Maeda1,ISHI}
It was shown\cite{ISHI} that there are two exponentially expanding solutions,
which may be called generalized de Sitter solutions since the size of the
internal space depends on time (otherwise there is no solution of this type).
In both solutions, the external space inflates, while the internal space
shrinks exponentially. (There are also two time-reversed solutions, i.e. the
external space shrinks exponentially but the internal space inflates.)
One solution is stable and the other is unstable. Since the present
universe is not in the phase of de Sitter expansion with this energy scale,
we cannot use the stable solution for a realistic universe. If we adopt
the unstable solution, on the other hand, we may not find sufficient
inflation unless we fine-tune the initial values.

However, these powers of scalar curvature or Lovelock gravity are not
the types of corrections arising in type II superstrings or M-theory.
In particular, it is known that the coefficient
of the GB terms vanishes and the first higher order corrections
start with $R^4$ terms (one is the fourth order Lovelock gravity and the
other contains higher derivatives).\cite{TBB}
Probably due to the complicated nature of these corrections, this case was
not studied and so it is important to examine how these corrections in the
fundamental theories modify the above cosmological models and
whether we can get interesting cosmological scenario with large e-foldings.
The solutions to the vacuum Einstein equations with these higher
order corrections are studied.\cite{MO}

Before proceeding to our discussions of these corrected theories,
we note that the realization of de Sitter solutions in superstring theories
is notoriously difficult. As the no-go theorem indicates, accelerating
expansion, which is typical of de Sitter solutions, cannot be obtained
at the stationary minima in superstring or supergravity theories.
Various attempts to avoid this theorem are made by considering warped
solutions for noncompact gauged supergravities,\cite{GH}
by including hypermultiplet in gauged supergravities,\cite{desitter1}
or by including nonperturbative effects in heterotic M-theory.\cite{desitter2}
Note that most of these attempts require corrections to the lowest
supergravity theories. We shall show that the quantum corrections
in M-theory and type II superstrings mentioned above also
give generalized de Sitter solutions without adding anything special.
One thing worth mentioning is that this class of solutions do not exist
if we keep the size of the internal space constant, reminiscent
of the no-go theorem.
Moreover this de Sitter phase decays after some period
of inflation because of the small instability in the solutions.
This is a very interesting mechanism to end the inflation naturally.

We consider the low-energy effective action for the superstrings and/or
M-theory:
\bea
S &=& S_{\rm EH} + S_{\rm GB} + S_4 + S_S,
\label{totaction}
\ena
where
\bea
S_{\rm EH} &=& \frac{1}{2\kappa_D^2} \int d^D x \sqrt{-g} R,\\
\label{gb}
S_{\rm GB} &=& \frac{\a}{2\kappa_{D}^2} \int d^{D} x \sqrt{-g}\;  G,\\
S_4 &=&  \frac{\b}{2\kappa_{D}^2}\int d^D x
 \sqrt{-g} \tilde{E}_8, \\
S_S &=&  \frac{\c}{2\kappa_{D}^2}\int d^D x \sqrt{-g} \tilde{J}_0\,.
\label{4th}
\ena
\bea
\tilde{E}_8&=&-{1\over 2^4 \times 3!}
\e^{\a\b\c\mu_1 \nu_1 \ldots \mu_4 \nu_4}
\e_{\a\b\c\rho_1 \sigma_1 \ldots \rho_4 \sigma_4} R^{\rho_1\sigma_1}
{}_{\mu_1 \nu_1} \cdots R^{\rho_4 \sigma_4}{}_{\mu_4 \nu_4}\,,
\\
\tilde{J}_0&=&
R^{\lambda\mu\nu\kappa}R_{\a\mu\nu\b}R_{\lambda}{}^{\rho\sigma\a}
R^\b{}_{\rho\sigma\kappa}
+\frac12 R^{\lambda\kappa\mu\nu}R_{\a\b\mu\nu}R_{\lambda}{}^{\rho\sigma\a}
R^\b{}_{\rho\sigma\kappa}.
\ena
Here we have dropped contributions from forms, $\kappa_D^2$ is
a $D$-dimensional gravitational constant.
For the heterotic string, the leading correction is given by the
GB terms with the coefficient:\cite{Be,hetero}
\bea
\a=\frac{1}{8}\alpha',
\label{hetero}
\ena
if we keep the dilaton constant, where $\alpha'$ is the Regge slope parameter.
For the M-theory in 11 dimensions, the coefficient for the
GB terms $\a$ vanishes, so we should consider forth order
terms with the coefficients:\cite{TBB}
\bea
\beta = - {\kappa_{11}^2 ~T_2\over 3^2\times 2^{9} \times (2\pi)^4} ,\qquad
\gamma = - {\kappa_{11}^2 ~ T_2\over 3 \times 2^{4}\times (2\pi)^4} \,,
\label{m-theoryc}
\ena
where $T_2=({2\pi^2 /\kappa_{11}^2})^{1/3}$ is the membrane tension.
Type II superstring has the same couplings in 10 dimensions,
so we can discuss this case if we keep the dilaton field constant,
but we consider 11D theory in this paper.
Here we should note that contributions of the Ricci tensor $R_{\mu\nu}$
and scalar curvature $R$ are not included in the fourth-order
corrections~(\ref{4th}) because these terms are not uniquely fixed.

Since we are interested in a cosmological time-dependent solution,
we take the metric of our spacetime as
\bea
ds_{11}^2 &=& -e^{2u_0(t)} dt^2 + e^{2u_1(t)} \sum_{i=1}^3 (dx^i)^2
+ e^{2u_2(t)} \sum_{a=5}^{11} (dy^a)^2\,,
\label{met1}
\ena
where we assume that the external 3-space and the internal 7-space
are flat. Taking variation of the action with respect to $u_0$,
$u_1$, and $u_2$, we obtain three basic equations, whose explicit forms
are summarized in Appendix.

\subsection{Generalized de Sitter Solutions}

In cosmology, de Sitter inflationary expanding spacetime is the most
interesting solution in the early universe.
We study such solutions to the M-theory with higher order corrections
with $\a=0$ and other coefficients given in \p{m-theoryc}.\cite{MO}
We can also consider EH theory with corrections
of GB term, but these solutions are already known.\cite{ISHI}
Extensive study of the generalized de Sitter and power-law solutions
is also made for the external and internal spaces with nontrivial
curvatures and more inflationary solutions are found,\cite{MO1}
but here we discuss only flat spaces.

Assuming the metric form of a generalized de Sitter spacetime as
\bea
u_0=0 ,\,\,u_1=\mu t ,\,\,u_2=\nu t ,\,\,
\ena
where $\mu$ and $\nu$ are some constants, we obtain three algebraic
equations from the general field equations given in Appendix:
\bea
\label{1}
&& \mu^2 + 7\mu\nu + 7\nu^2
 + 20160 \b  \mu \nu^5
\Big[ 7 \mu^2 + 7 \mu \nu +  \nu^2 \Big]  \nn
&& -7 \c \Big[ 12 \mu^8 + 7\mu^2 \nu^2 (\mu^2+\nu^2+\mu\nu )^2
+168 \nu^8 +7 \mu^4 \nu^2(2\mu+\nu)^2  +21\mu^2 \nu^4(\mu+2\nu)^2  \Big]
\nn && + 4\c (3\mu+7\nu)\Big[ 6 \mu^7 + 42 \nu^7 + 7\mu^2 \nu^2 (\mu+\nu)
(\mu^2+\nu^2+\mu\nu ) \Big] = 0, \\
\label{2}
&& 3 \mu^2 + 14\mu\nu + 28 \nu^2  + 20160 \b \nu^5\Big[ 6 \mu^3 + 24
\mu^2 \nu + 14 \mu\nu^2+  \nu^3 \Big]
\nn
&& +3 \c \Big[ 12 \mu^8 + 7\mu^2 \nu^2
(\mu^2+\nu^2+\mu\nu )^2
 +  168 \nu^8 + 7\mu^4 \nu^2(2\mu+\nu)^2  + 21\mu^2 \nu^4(\mu+2\nu)^2
\Big]
\nn
&& -2 \c \mu (3\mu+7\nu)\Big[ 48 \mu^6 + 7 \nu^2 (3\mu^2  +2\mu\nu +\nu^2)
(\mu^2+\nu^2+\mu\nu ) \nn
&&~~~~~~~~~~~~~~~~~~~~ +14\mu^2 \nu^2(2\mu+\nu)(3\mu+\nu)  +
42 \nu^4(\mu+2\nu)(\mu+\nu) \Big] \nn
&& +2\c \mu (3\mu+7\nu)^2\Big[ 12 \mu^5+7 \nu^2(\mu+\nu)  (\mu^2+\nu^2
+\mu\nu )
\Big] = 0, \\
\label{3}
&& 2\mu^2 + 6\mu\nu + 7 \nu^2 + 2880\b \mu\nu^4 \Big[
15\mu^3 + 46 \mu^2 \nu   +38\mu\nu^2 + 6 \nu^3
\Big] \nn
&& + \c \Big[ 12 \mu^8 + 7 \mu^2 \nu^2(\mu^2+\nu^2+\mu\nu )^2  +168
\nu^8 +7\mu^4 \nu^2(2\mu+\nu)^2    +21\mu^2 \nu^4(\mu+2\nu)^2  \Big]
\nn
&& -2 \c \nu (3\mu+7\nu)\Big[ 96 \nu^6  +
\mu^2   (\mu^2+2\mu\nu+3\nu^2)(\mu^2+\nu^2+\mu\nu )
\nn &&
~~~~~~~~~~~~~~~~~~~~+2\mu^4  (2\mu+\nu)(\mu+\nu) +6\mu^2
\nu^2(\mu+2\nu)(\mu+3\nu) \Big]
\nn && +2\c\nu (3\mu+7\nu)^2\Big[ 12 \nu^5+\mu^2  (\mu+\nu) (\mu^2+\nu^2
+\mu\nu )
\Big]=0 \,.
\ena

Since these equations are very complicated, we have solved them numerically.
Rescaling $\b$, $\c$, $\mu$ and $\nu$ as
\bea
\tilde{\b}=\b /|\c| \, \,,
\tilde{\c}=\c /|\c| ~(=1 ~{\rm or}~ -1)  \, \,,
\tilde{\mu}= \mu |\c|^{1/6}  \,, \,{\rm and}\,\,
\tilde{\nu}= \nu |\c|^{1/6}\,,
\ena
we can always set $\c$ to $-1$.
We also have to rescale time coordinate as $\tilde{t}=|\c|^{-1/6} t$.
The typical dynamical time scale is then given by $|\c|^{1/6}\sim
0.181818 m_{11}^{-1}$, where $m_{11}=\kappa_{11}^{-2/9}$ is
the fundamental Planck scale. After this scaling, we have only one
free parameter $\tilde{\b}$.
We find that M-theory
($\tilde{\c}=-1,\tilde{\b}=\tilde{\b}_S=-1/(3\times 2^5) \approx -0.0104167$)
has three solutions
\bea
\label{oursol2}
{\rm N}_2 (\tilde{\mu}_2,\tilde{\nu}_2) &=& (0.40731,0.40731),\nn
{\rm N}_3 (\tilde{\mu}_3,\tilde{\nu}_3) &=& (0.79683,0.10793),\\
{\rm N}_4 (\tilde{\mu}_4,\tilde{\nu}_4) &=& (0.55570,0.34253),\nonumber
\ena
and their time-reversed solutions N$'_{i}(\tilde{\mu}'_i,\tilde{\nu}'_i)$
($i=2 \sim 4$) obtained by $(\tilde{\mu}'_i,\tilde{\nu}'_i)
=-(\tilde{\mu}_i,\tilde{\nu}_i)$.
Other possible solutions for various values of $\b$ and $\c$ are
given in Ref.~\refcite{MO}.

\subsection{Stability}

Since the solutions obtained above correspond to fixed points in our dynamical
system, we can see which solutions are more generic by looking at
their stabilities. A linear perturbation analysis around those fixed points
was carried out.\cite{MO} Setting
\bea
\frac{d{u}_1}{d\tilde{t}}=\tilde{\mu}_i +A_i e^{\sigma \tilde{t}},\quad
\frac{d{u}_2}{d\tilde{t}}=\tilde{\nu}_i +B_i e^{\sigma \tilde{t}},
\ena
where
$|A_i|,|B_i|\ll 1$, the perturbation equations are written down.
There are five modes ($\sigma=\sigma_a^{(i)},~a=1,2,\cdots, 5$).
The M-theory has three solutions~\p{oursol2}.
Two solutions (${\rm N}_2$ and ${\rm N}_3$) have four stable and
one unstable modes. The third solution (${\rm N}_4$) has five stable
modes, which means that this solution is stable against linear perturbations.

Which of these solutions is most desirable?
We want a solution to be rather generic which requires some sort of stability
with natural mechanism of ending inflation.
This would be achieved if the solution contains many stable as well as
tiny unstable modes. Then the spacetime may first approach this solution
for wide range of initial conditions and gradually leave it due to the
tiny instability, recovering the present Friedmann universe, where we
expect the higher order terms become irrelevant. The solution ${\rm N}_3$
may give one possible candidate for such a model.

\subsection{A Scenario for Large Extra Dimensions}

For the interesting solution ${\rm N}_3 (\tilde{\mu}_3,\tilde{\nu}_3)
=(0.79683,0.10793)$ with $\b=\b_{S}$, the scale factor of the external
space expands as $e^{0.79683\tilde{t}}$. A successful inflation
(resolution of flatness and horizon problems) requires at least 60 e-foldings.
We assume that inflation ends after 60 e-foldings, i.e.
$0.79683  \tilde{t}_{\rm end}\approx 60$ due to the unstable mode.
During inflation the internal space also expands exponentially and its
scale becomes $e^{0.10793 \tilde{t}_{\rm end}}\approx 4000$ times larger
than the initial scale length, which we assume to be the 11D Planck length
($m_{11}^{-1}$). The present radius of extra dimensions is then estimated as
$R_0 \sim 4000 m_{11}^{-1}$. This gives us a model of large extra
dimensions.\cite{arkani} The 4D Planck mass is given by
\bea
m_4^2\sim R_0^7 m_{11}^9 \sim  1.6\times 10^{25} m_{11}^2.
\label{m4}
\ena
We then find
\bea
m_{11}\sim 2.5 \times 10^{-13} m_{4} \sim 600 {\rm TeV} \,.
\ena
This is our fundamental energy scale.
The present scale of extra dimensions is $4000 m_{11}^{-1}\sim 7$
TeV$^{-1}$, which could be observed in the accelerators of next generation.

By putting the argument in a opposite way, we find that this solution
gives a natural explanation why the e-folding becomes of the order 60.
First suppose that the e-folding of inflation is $N$. The 3-space expands as
$e^N=e^{\tilde{\mu}\tilde{t}_{\rm end}}$, while the internal space
becomes $e^{\tilde{\nu}\tilde{t}_{\rm end}}$ times larger. It follows from
Eq.~\p{m4} that
\bea
m_{11}\sim e^{-{7\nu \over 2\mu}N} m_4\,.
\label{extrap}
\ena
Since $m_{11}\gsim 1$ TeV from the present experiments, we have a
constraint on the e-folding as $N\lsim 10\mu/\nu$. Then if we have TeV
gravity and $\mu \gsim 6\nu (> 0)$, this naturally explain why the
e-folding of inflation is about 60 and but not so large.
Here it is important to note that the solution ${\rm N}_3$ with $\gamma<0$
gives $5.72<\mu/\nu<10.22$ (corresponding to $57\lsim N\lsim 102$)
for any value of $\beta$.

Although the above solution ${\rm N}_3$ has one unstable mode, its
eigenvalue is of the same order of magnitude as other eigenvalues of
stable modes and is a little too large to give enough expansion.
It is desirable that the eigenvalue of the unstable mode is much smaller
than those of other four stable modes for obtaining large e-folding.
The question is if this is possible within this framework of M-theory.

Our starting Lagrangian has ambiguity in the fourth-order correction
terms, which are fixed up to the Ricci and scalar curvature tensors.
The effect of the ambiguous Ricci and scalar curvature tensors will modify
the basic equations. Assuming that this can be taken into account by changing
the value of the coefficient $\tilde \b$, such a desirable solution is
searched for. An interesting solution is found for $\tilde{\b}=-0.2025$,
for which four modes are stable and the eigenvalue of one unstable mode
is very small, i.e. $\sigma^{(3)}_5=0.01217$. Similar estimate of the
size of the extra dimensions then gives again a few TeV, resulting in
TeV gravity.

There are several bounds on the allowed size of the extra dimensions from
supernova 1987A cooling,\cite{ADD,CP,BHK1} from graviton decay into diffuse
gamma radiation,\cite{HS2} or from collider experiments.\cite{GRW}
All these indicate that the size of the extra dimensions should be
greater than a few TeV, which is barely of the same order of magnitude
as the above results. So the extra dimensions might be discovered
in the next generation of accelerators or further study of astronomical
constraints.

If we consider external and internal spaces with nontrivial curvatures,
it has been found that there are several more interesting solutions which
can give inflationary solutions in the Einstein frame in four
dimensions.\cite{MO1} There not only exact solutions to the basic
equations with higher order corrections but also asymptotic past and
future solutions are exhausted for generalized de Sitter and power-law
solutions. The past asymptotic solutions are useful for describing
inflations at early time whereas the future one for understanding
accelerating cosmologies at the present time.
Cosmological models based on these are under study\cite{AMO}
and this approach is expected to lead to further interesting models
with large e-foldings for the inflation at early epoch of our universe
as well as the present accelerating universe.

Whatever modifications we make to improve some problems or other
in these solutions, the emerging common feature of these solutions
is that the size of the extra dimensions are large and the value of
eleven-dimensional Planck mass is small. What do they mean? They mean
that when we make experiments beyond the size of extra dimensions or
energy of a few TeV, we begin to see new dimensions open up and strong
gravitational effects in higher dimensions set in. There we may have
effects of strong gravitational fields and black holes, and have to
take full quantum gravity effects into account to understand phenomena
at this energy scale.\cite{GK,DL} Or we can probe how the quantum
gravity behaves there. This is a very interesting possibility.

Although we find a successful exponential expansion and its natural
end in this simple setting without introducing any artificial objects into
M-theory, this is not enough for a successful inflation.
We need a reheating mechanism and have to create a density fluctuation
as a seed of cosmic structure.
A possible mechanism of reheating is the gravitational particle
creation,\cite{reheating} because the background spacetime is time
dependent and this might have some oscillation when the internal space
settles down to static one which is required to explain our present
universe. As for a density perturbation, our model may not give a
good scenario because our energy scale is $|\gamma|^{-1/6} \sim 5 m_{11}
\sim 4$ TeV. We have to use other mechanism for density perturbations
such as a curvaton model.\cite{curvaton}
These are the problems left for future study.

\section{Other approaches to accelerating universe}

The no-go theorem states that no smooth classical compactification of
M-theory leads to an accelerating universe. However there are several
assumptions in deriving this claim as discussed in Sec.~\ref{avoid},
so there is a possibility to avoid this theorem if any one of them is
violated. The approaches using time-dependent internal spaces and higher
order corrections are already discussed. Here we summarize several
other attempts in this direction.

It has been argued that Type II$^*$ string theories, which are obtained from
time-like duality, have Euclidean branes  which have de Sitter space
in the near-horizon limit.\cite{Hull} A possible problem is
how to make sense of the Type II$^*$ theories because their low-energy
effective theories contain higher derivative terms and wrong sign
kinetic terms arising from the timelike duality.
This is similar to what is called ``phantom cosmology.''\cite{phantom,Faraoni}
The recent idea to combine such ghost and higher derivative terms to
stabilize a theory could be put in this context.\cite{ACL}

A compactification of M-theory on a classically singular manifold like
a line of finite length is another way to avoid the theorem due to the
existence of the singularity in the manifold. By including
extra stringy states like wrapped branes, a smooth low-energy action was
derived despite the singularity, and it has been shown that exciting
such extra states leads to a period of accelerated expansion,\cite{JMS}
but it seems difficult to obtain enough e-foldings since the duration
of the expansion is typically short.

Asymmetric orientifold is also used to construct de Sitter vacua.\cite{Silv}
It was argued that it is possible to stabilize all the moduli at a
positive extremum of the potential by turning on fluxes in such models.
This shares many features with the recent discussions of flux
compactification.\cite{KKLT}

Many light moduli fields are generated in these string compactifications.
The problem is that these fields are difficult to stabilize and tend to
be runaway which leads to decompactification to 10 dimensions.
In the flux compactification, it is argued that they can be stabilized
by turning on fluxes on the compactification manifolds. The argument
is that nonperturbative effects generate potentials which stabilize such
a runaway at supersymmetric extrema. Similar discussion is given
in Ref.~\refcite{Biswas} using dimensionally reduced theories with
scalars and antisymmetric tensors. But this is not enough to
get a de Sitter vacuum because all supersymmetric vacua are in anti-de Sitter.
We have to break supersymmetry but simple breaking of supersymmetry
is not good enough since it will lead to big cosmological constant. It was
argued that by adding an anti-D3 brane, supersymmetry can be broken in a
controllable way and we can get metastable de Sitter vacuum.\cite{KKLT,Kachru}
However constructing an explicit example of this sort is difficult
and no successful scenario has been given. There are lots of efforts
to improve this and it is possible that more promising scenario will be given.
Possible modifications are proposed by considering Dirac-Born-Infeld action
with higher-derivative terms which enforce the slow roll of the scalar
fields.\cite{ST}

Variants of the higher order corrections discussed above were also
considered with negative power of the scalar curvature and/or mixed
system with positive powers for the purpose of generating
the present accelerating cosmologies and inflation.\cite{CDT,NOD,GZB}
Though this is certainly possible in this modified theory, this modification
gives rise to the change of the gravitational law at low curvature and
simple models are in conflict with solar system tests of gravity.\cite{DK}
Ricci curvature squared is also considered with the results of accelerating
cosmology.\cite{ABF}
Another common problem with these proposals is that the origin of
such terms is not clear.

Some other attempts to derive accelerating cosmologies in higher-dimensional
Kaluza-Klein theories includes Refs.~\refcite{Mohammedi,Darabi} with
exotic matter. Also systematic methods to find time-dependent
solutions in these theories are developed using group theoretic
method\cite{Fre} or extending the solution space such that the cosmological
solutions are geodesic motion in that space,\cite{TW2} and they may be
useful to find interesting solutions.

Theories with large number of scalar fields are also considered.\cite{BHK2}
This is motivated from the fact that in superstring theories there are
a large number of scalar fields. Consistency with cosmic microwave
background data suggests that the theory should have some symmetry to
avoid excessive isocurvature perturbations. Thus theories with $SO(N)$
symmetry is studied and it is found that the spectrum of fluctuations
are different from single inflaton case.

\section{Conclusions}

In this paper we have reviewed recent progress in the accelerating
cosmologies and inflationary solutions in superstrings or supergravities.
The main focus is our time-dependent solutions based on S-brane solutions
in M-theory and superstrings with and without higher order corrections,
but we have also mentioned several other approaches discussed actively.
There are so many directions and literature that we cannot cover all of them.
Since the superstring theories or M-theory is supposed to give the
quantum theory of gravity, any attempts to give inflation and the present
accelerating expansion of our universe should eventually be made in
these theories. Then how to avoid the no-go theorem is the main
first step to get such solutions.

Our approach using the S-brane solutions
is based on the violation of the assumption of time-independence of the
size of internal space. We have also shown that the higher order corrections
can produce interesting inflationary solutions. The common feature obtained
in these solutions is that the size of extra dimensions are rather large
of the order of TeV. Future experiments probing this energy scale might show
anomalous behaviors related to strong gravitational interactions.

The solutions reviewed here have the virtue that they are the solutions
of simple systems without introducing branes or other objects.
Any scenario involving such objects would need explanations why such
configurations are prepared, but it may be difficult to explain it naturally.
The solutions discussed here are natural in the sense not only that they
are simply the solutions to the field equations of the fundamental theories
without no other input, but also that, as we have shown for solutions in
M-theory with fourth-order corrections, general solutions tend to converge
the (quasi)stable solutions for wide range of initial conditions.
This class of solutions are also interesting in that they have inherent
mechanism of ending inflation; they eventually decays due to the tiny
instability.

These are the desirable properties, but there also remain some problems
(not difficulties) to be studied further. We need a mechanism of keeping
the size of the internal space for the S-brane solutions.
How to derive reheating and create density fluctuation are the problems
that should be studied. We hope that this review may stimulate further
study of these approaches to accelerating cosmologies and inflation
in the context of superstring theories and M-theory.

\section*{Acknowledgements}

I would like to thank C. M. Chen, P. M. Ho, I. Neupane, J. E. Wang
and K. Maeda for fruitful collaboration and discussions.
Thanks are also due to P. K. Townsend and M. N. R. Wohlfarth for discussions.
The work was supported in part by the Grant-in-Aid for Scientific Research
Fund No. 16540250.

\appendix

\section{Explicit form of the equations}
\label{a:ein}

Here we give general field equations in the theories with higher order
corrections for the metric similar to~\p{met1}, but slightly generalized to
$p$-dimensional external and $q$-dimensional internal spaces with
nontrivial curvatures whose signatures are denoted by $\s_p$ and $\s_q$,
respectively.
The action with higher order corrections~\p{totaction} takes the form\cite{MO2}
\bea
&& S_{\rm EH}
= \frac{1}{16\pi G} \int d^{p+1}x \;
e^{u_0 + pu_1+(q-2) u_2} \Big[ p_1 \s_p e^{2u_2-2u_1} + q_1 \s_q \nn
&& \hs{30}
-\; \Big\{ p_1 \dot u_1^2 + q_1 \dot u_2^2 +2pq \dot u_1 \dot u_2 \Big\}
 e^{-2u_0 +2u_2} \Big], \\
&& S_{\rm GB}
= \frac{\a}{16\pi G} \int d^{p+1}x \;
e^{-u_0 +pu_1+(q-2) u_2} \Big[ p_3 \s_p^2 e^{2u_0 -4u_1 + 2u_2}
+ 2p_1 q_1 \s_p \sigma_q e^{2u_0 -2u_1} \nn
&& \hs{20}
+ q_3 \sigma_q^2 e^{2u_0 -2u_2}
- 2q_1 \s_q \Big\{ p_1 \dot u_1^2 + 2p(q-2) \dot u_1 \dot u_2
+(q-2)_3 \dot u_2^2 \Big\} \nn
&& \hs{20}
-\; 2p_1 \s_p \Big\{ (p-2)_3 \dot u_1^2 + 2q(p-2) \dot u_2 \dot u_1
+ q_1 \dot u_2^2 \Big\} e^{2u_2-2u_1} \nn
&& \hs{10}
-\; \frac{1}{3} \Big\{ p_3 \dot u_1^4 +4qp_2 \dot u_1^3
\dot u_2 + 6p_1 q_1 \dot u_1^2 \dot u_2^2
+ 4pq_2 \dot u_1 \dot u_2^3
+ q_3 \dot u_2^4 \Big\} e^{-2u_0 +2u_2} \Big],
\label{action:GB}
\\
&& S_4 = -\frac{\b/7}{16\pi G}
\int d^{1+p} x\;
\sum_{k=0}^8 \frac{8!\; p!\; q!\; \dot u_1^{8-k} \dot u_2^k\;
 e^{-7 u_0 +pu_1+qu_2}}{k!\; (8-k)!\; (p-8+k)!\; (q-k)!} , \\
&& S_S = \frac{1}{16\pi G} \int d^{1+p} x\; \c\; e^{-7u_0 + pu_1+q u_2} \Big[
p_1 X^2 \{ X+2(\dot u_1^2+\s_p e^{2u_0-2u_1}) \}^2 \nn
&& \hs{10}
+\; q_1 Y^2 \{ Y+2(\dot u_2^2+\s_q e^{2u_0-2u_2}) \}^2
+\; 2pq \{ XY+(X+Y) \dot u_1 \dot u_2 \}^2 \nn
&& \hs{10}
+\; 3p_2 (\dot u_1^2+\s_p e^{2u_0-2u_1})^4
+\; 3q_2 (\dot u_2^2+\s_q e^{2u_0-2u_2})^4 \nn
&& \hs{10}
+\; p_1 q \{ \dot u_1 \dot u_2 + 2(\dot u_1^2+\s_p e^{2u_0-2u_1})\}^2
\dot u_1^2 \dot u_2^2 \nn
&& \hs{10}
+\; pq_1 \{ \dot u_1 \dot u_2 + 2(\dot u_2^2+\s_q e^{2u_0-2u_2})\}^2
\dot u_1^2 \dot u_2^2 \Big],
\label{action:r4}
\ena
where $S_4$ is given only for $\s_p=\s_q=0$ case but others are
for general case, and we use the following notation throughout this paper:
\bea
(p-m)_n&\equiv& (p-m)(p-m-1)(p-m-2)\cdots (p-n),
\nonumber \\
(q-m)_n&\equiv&(q-m)(q-m-1)(q-m-2)\cdots (q-n),
\nonumber \\
A_p&\equiv&\dot{u}_1^2+\sigma_p e^{2(u_0-u_1)}, \quad
A_q \equiv \dot{u}_2^2+\sigma_q e^{2(u_0-u_2)},\nn
X &\equiv& \ddot u_1 - \dot u_0 \dot u_1 +\dot u_1^2, \quad
Y \equiv \ddot u_2 - \dot u_0 \dot u_2 +\dot u_2^2.
\label{xy}
\ena

{}From the variation of the total action~\p{totaction} with respect to
$u_0, u_1$ and $u_2$, we find three basic field equations for general
case with nonvanishing curvature:\cite{MO1}
\bea
\label{eq1}
&&
F\equiv \sum_{n=1}^4 F_n+F_S=0\,,
\\
\label{eq2}
&&
F^{(p)}\equiv \sum_{n=1}^4
f_n^{(p)}+\left(\sum_{n=1}^4
g_n^{(p)}\right)X+\left(\sum_{n=1}^4
h_n^{(p)}\right)Y+F_S^{(p)}=0\,,
\\
\label{eq3}
&&
F^{(q)}\equiv \sum_{n=1}^4
f_n^{(q)}+\left(\sum_{n=1}^4
g_n^{(q)}\right)Y+\left(\sum_{n=1}^4
h_n^{(q)}\right)X+F_S^{(q)}=0\,,
\ena
where each terms are summarized below.\\
{\bf (1) EH action}
\bea
\label{eh1}
F_1&=& e^{-u_0}\left[p_1 A_p+q_1 A_q+2pq\dot{u}_1\dot{u}_2\right],
\\
f_1^{(p)}&=& e^{-u_0}\left[
(p-1)_2A_p+q_1 A_q+2(p-1)q\dot{u}_1\dot{u}_2\right],
\\
f_1^{(q)}&=& e^{-u_0}\left[p_1 A_p+
(q-1)_2A_q+2p(q-1)\dot{u}_1\dot{u}_2\right],
\\
g_1^{(p)}&=&2(p-1) e^{-u_0},
\\
g_1^{(q)}&=&2(q-1) e^{-u_0},
\\
h_1^{(p)}&=&2q e^{-u_0},
\\
h_1^{(q)}&=&2p e^{-u_0}.
\ena
{\bf (2) GB action}
\bea
&& F_2 = \a e^{-3u_0} [ p_3 A_p^2+2p_1q_1 A_pA_q+q_3 A_q^2
+4\dot{u}_1\dot{u}_2(p_2 q A_p+p q_2 A_q)\nn
&& \hs{30}
+\; 4p_1q_1\dot{u}_1^2\dot{u}_2^2 ],
\\
&& f_2^{(p)}\; =\; \a e^{-3u_0}\left[
(p-1)_4A_p^2+2(p-1)_2q_1 A_pA_q+q_3 A_q^2
\right.\nonumber \\
&& \hs{10}
\left.
+\; 4\dot{u}_1\dot{u}_2
\left((p-1)_3qA_p+(p-1)q_2
A_q\right)+4(p-1)_2q_1\dot{u}_1^2\dot{u}_2^2\right],
\\
&& f_2^{(q)}\; =\; \a e^{-3u_0}\left[
(q-1)_4A_q^2+2(q-1)_2p_1 A_pA_q+p_3 A_p^2
\right.\nonumber \\
&& \hs{10}
\left.
+\; 4\dot{u}_1\dot{u}_2
\left((q-1)_3pA_q+(q-1)p_2
A_p\right)+4(q-1)_2p_1\dot{u}_1^2\dot{u}_2^2\right],
\\
&& g_2^{(p)}\; =\; 4(p-1)\a e^{-3u_0}\left[
(p-2)_3A_p+q_1A_q+2(p-2)q\dot{u}_1\dot{u}_2\right],
\\
&& g_2^{(q)}\; =\; 4(q-1)\a e^{-3u_0}\left[
(q-2)_3A_q+p_1A_p+2(q-2)p\dot{u}_1\dot{u}_2\right],
\\
&& h_2^{(p)}\; =\; 4q\a e^{-3u_0}\left[
(p-1)_2A_p+(q-1)_2A_q+2(p-1)(q-1)\dot{u}_1\dot{u}_2\right],
\\
&& h_2^{(q)}\; =\; 4p\a e^{-3u_0}\left[
(q-1)_2A_q+(p-1)_2A_p+2(p-1)(q-1)\dot{u}_1\dot{u}_2\right].
\label{gbl}
\ena
{\bf (3) Lovelock action}
\bea
F_4&=&
\b e^{-7u_0} \left[
p_7 A_p^4+4p_5q_1 A_p^3A_q+6p_3q_3A_p^2A_q^2
+4p_1q_5 A_pA_q^3+q_7 A_q^4
\right. \nn
&&
+8\dot{u}_1\dot{u}_2
(p_6 q A_p^3+3p_4q_2A_p^2A_q+3p_2q_4A_pA_q^2 +p q_6 A_q^3)
+24\dot{u}_1^2\dot{u}_2^2(p_5q_1A_p^2\nn
&&\left.
+2p_3q_3A_pA_q+p_1q_5A_q^2)
+32\dot{u}_1^3\dot{u}_2^3(p_4q_2A_p+p_2q_4A_q)
+16p_3q_3\dot{u}_1^4\dot{u}_2^4 \right],
\\
f_4^{(p)}&=&\b e^{-7u_0}\left[ (p-1)_8 A_p^4+4(p-1)_6q_1
A_p^3A_q+6(p-1)_4q_3A_p^2A_q^2
\right.\nn
&&
+4(p-1)_2q_5A_pA_q^3+q_7 A_q^4
+8\dot{u}_1\dot{u}_2 \left((p-1)_7qA_p^3+3(p-1)_5q_2A_p^2A_q \right. \nn
&&
\left.
+3(p-1)_3q_4A_pA_q^2+(p-1)q_6 A_q^3\right)
+24\dot{u}_1^2\dot{u}_2^2\left((p-1)_6q_1A_p^2 \right.\nn
&& \left.
+2(p-1)_4q_3A_pA_q +(p-1)_2q_5A_q^2\right)
\nonumber \\
&&
\left.
+32\dot{u}_1^3\dot{u}_2^3\left( (p-1)_5q_2A_p+(p-1)_3q_4 A_q\right)
+16(p-1)_4q_3\dot{u}_1^4\dot{u}_2^4\right],
\\
f_4^{(q)}&=&\b e^{-7u_0}\left[ (q-1)_8 A_q^4+4(q-1)_6p_1
A_q^3A_p+6(q-1)_4p_3A_p^2A_q^2
\right.\nn
&&
+4(q-1)_2p_5A_qA_p^3
+p_7 A_p^4+8\dot{u}_1\dot{u}_2 \left((q-1)_7pA_q^3+3(q-1)_5p_2A_q^2A_p
\right.
\nonumber \\
&&\left.
+3(q-1)_3p_4A_qA_p^2+(q-1)p_6 A_p^3\right)
+24\dot{u}_1^2\dot{u}_2^2\left((q-1)_6p_1A_q^2 \right.\nn
&&
\left.
+2(q-1)_4p_3A_pA_q
+(q-1)_2p_5A_p^2\right) \nn
&& \left.
+32\dot{u}_1^3\dot{u}_2^3\left( (q-1)_5p_2A_q+(q-1)_3p_4 A_p\right)
+16(q-1)_4p_3\dot{u}_1^4\dot{u}_2^4 \right],
\\
g_4^{(p)}&=&8(p-1)\b e^{-7u_0}\left[
(p-2)_7A_p^3+3(p-2)_5q_1A_p^2A_q+3(p-2)_3q_3A_pA_q^2+q_5A_q^3
\right.\nonumber\\
&&
\left.
+6\dot{u}_1\dot{u}_2\left( (p-2)_6qA_p^2+2(p-2)_4q_2A_pA_q
+(p-2)q_4A_q^2\right)\right.
\nonumber\\
&&
\left.
+12\dot{u}_1^2\dot{u}_2^2\left( (p-2)_5q_1A_p+(p-2)_3q_3A_q\right)
+8(p-2)_4q_2\dot{u}_1^3\dot{u}_2^3\right],
\\
g_4^{(q)}&=&8(q-1)\b e^{-7u_0}\left[
(q-2)_7A_q^3+3(q-2)_5p_1A_q^2A_p+3(q-2)_3p_3A_qA_p^2+p_5A_p^3
\right.\nonumber\\
&&
\left.
+6\dot{u}_1\dot{u}_2\left( (q-2)_6pA_q^2+2(q-2)_4p_2A_pA_q
+(q-2)p_4A_p^2\right)\right.
\nonumber\\
&&
\left.
+12\dot{u}_1^2\dot{u}_2^2\left( (q-2)_5p_1A_q+(q-2)_3p_3A_p\right)
+8(q-2)_4p_2\dot{u}_1^3\dot{u}_2^3\right],
\\
h_4^{(p)}&=&8q\b e^{-7u_0}\left[
(p-1)_6A_p^3+3(p-1)_4(q-1)_2A_p^2A_q+3(p-1)_2(q-1)_4A_pA_q^2
\right.
\nonumber\\
&&
+(q-1)_6A_q^3
+6\dot{u}_1\dot{u}_2\left( (p-1)_5(q-1)A_p^2+2(p-1)_3(q-1)_3A_pA_q
\right. \nn
&&
\left.
+(p-1)(q-1)_5A_q^2\right)
+12\dot{u}_1^2\dot{u}_2^2\left( (p-1)_4(q-1)_2A_p+(p-1)_2(q-1)_4A_q\right)
\nn
&& \left.
+8(p-1)_3(q-1)_3\dot{u}_1^3\dot{u}_2^3\right],
\\
h_4^{(q)}&=&8p\b e^{-7u_0}\left[
(q-1)_6A_q^3+3(q-1)_4(p-1)_2A_q^2A_p+3(q-1)_2(p-1)_4A_qA_p^2
\right.
\nn
&&
+(p-1)_6A_p^3
+6\dot{u}_1\dot{u}_2\left( (q-1)_5(p-1)A_q^2+2(p-1)_3(q-1)_3A_pA_q
\right. \nn
&&
\left.
+(q-1)(p-1)_5A_p^2\right)
+12\dot{u}_1^2\dot{u}_2^2\left( (q-1)_4(p-1)_2A_q+(q-1)_2(p-1)_4A_p\right)
\nn
&&  \left.
+8(p-1)_3(q-1)_3\dot{u}_1^3\dot{u}_2^3\right].
\ena
{\bf (4) $S_S$ action}
\bea
F_S &=& \c e^{-pu_1-qu_2} \Big[-7L_4+2\sigma_p e^{2(u_0-u_1)}{\partial
L_4\over \partial A_p}+2\sigma_q e^{2(u_0-u_2)}{\partial L_4\over
\partial A_q} \nn
&& \hs{20} +{d\over dt}\left(\dot{u}_1{\partial L_4\over
\partial X} +\dot{u}_2{\partial L_4\over \partial Y}\right)\Big]\,,
\\
pF_S^{(p)} &=& \c e^{-pu_1-qu_2} \left[
pL_4-2\sigma_p e^{2(u_0-u_1)}{\partial L_4\over \partial A_p} \right. \nn
&& \hs{5}\left.
+\; {d\over dt}\left((\dot{u}_0-2\dot{u}_1){\partial L_4\over \partial
X}-2\dot{u}_1{\partial L_4\over \partial A_p}-{\partial L_4\over \partial
\dot{u}_1}\right) + {d^2\over dt^2}\left({\partial L_4\over
 \partial X}\right) \right]\,,
\\
qF_S^{(q)} &=& \c e^{-pu_1-qu_2} \left[
qL_4-2\sigma_q e^{2(u_0-u_2)}{\partial L_4\over \partial A_q} \right. \nn
&& \hs{5} \left.
+{d\over dt}\left((\dot{u}_0-2\dot{u}_2){\partial L_4\over \partial
Y}-2\dot{u}_2{\partial L_4\over
\partial A_q}-{\partial L_4\over \partial \dot{u}_2}\right)
+ {d^2\over dt^2}\left({\partial L_4\over\partial Y}\right)
\right]
\,,
\ena
where
\bea
L_4 &=& e^{-7u_0+pu_1+qu_2}\left[p_1 X^2(X+2A_p)^2+q_1 Y^2(Y+2A_q)^2
+3p_2 A_p^4+3q_2 A_q^4
\right. \nn
&&
+\; 2pq(XY+(X+Y)\dot{u}_1\dot{u}_2)^2
+p_1 q\dot{u}_1^2\dot{u}_2^2(\dot{u}_1\dot{u}_2+2A_p)^2 \nn
&& \hs{20} \left.
+\; pq_1 \dot{u}_1^2\dot{u}_2^2(\dot{u}_1\dot{u}_2+2A_q)^2
\right],
\\
{\partial L_4\over\partial X} &=& 4pe^{-7u_0+pu_1+qu_2}\left[
(p-1)X(X+A_p)(X+2A_p) \right. \nn
&& \left. \hs{20} +\;q(Y+\dot{u}_1\dot{u}_2)(XY+(X+Y)\dot{u}_1\dot{u}_2)
\right], \\
{\partial L_4\over \partial Y} &=& 4qe^{-7u_0+pu_1+qu_2}\left[
(q-1)Y(Y+A_q)(Y+2A_q) \right. \nn
&& \hs{20} \left. +\;p(X+\dot{u}_1\dot{u}_2)(XY+(X+Y)\dot{u}_1\dot{u}_2)
\right], \\
{\partial L_4\over \partial A_p} &=& 4p_1 e^{-7u_0+pu_1+qu_2}\left[
X^2(X+2A_p)+3(p-2)A_p^3 \right.\nn
&& \hs{20}\left. +\;q\dot{u}_1^2\dot{u}_2^2(\dot{u}_1\dot{u}_2+2A_p)\right],
\\
{\partial L_4\over \partial A_q} &=& 4q_1 e^{-7u_0+pu_1+qu_2}\left[
Y^2(Y+2A_q)+3(q-2)A_q^3 \right.\nn
&& \hs{20}\left. +\;p\dot{u}_1^2\dot{u}_2^2(\dot{u}_1\dot{u}_2+2A_q)\right],
\\
{\partial L_4\over \partial\dot{u}_1} &=& 4pq~e^{-7u_0+pu_1+qu_2}~\dot{u}_2
\left[(X+Y)(XY+(X+Y)\dot{u}_1\dot{u}_2)
\right.
\nn
&&
+(p-1)\dot{u}_1\dot{u}_2
(\dot{u}_1\dot{u}_2+A_p)(\dot{u}_1\dot{u}_2+2A_p)\nn
&&
\left.
+(q-1)\dot{u}_1\dot{u}_2
(\dot{u}_1\dot{u}_2+A_q)(\dot{u}_1\dot{u}_2+2A_q)
\right],
\\
{\partial L_4\over\partial \dot{u}_2} &=&
4pq~e^{-7u_0+pu_1+qu_2}~\dot{u}_1\left[
(X+Y)(XY+(X+Y)\dot{u}_1\dot{u}_2)
\right.
\nn
&&
+(p-1)\dot{u}_1\dot{u}_2
(\dot{u}_1\dot{u}_2+A_p)(\dot{u}_1\dot{u}_2+2A_p) \nn
&&
\left.
+(q-1)\dot{u}_1\dot{u}_2
(\dot{u}_1\dot{u}_2+A_q)(\dot{u}_1\dot{u}_2+2A_q)
\right].
\ena

Since $u_0$ is a gauge freedom of time coordinate, we have three equations
\p{eq1} -- \p{eq3} for two variables $u_1$ and $u_2$. It looks like an
over-determinant system.
However, these three equations are not independent since
the following relation is valid:
\bea
\dot{F}+(p\dot{u}_1+q\dot{u}_2-\dot{u}_0) F=
p\dot{u}_1 F^{(p)}+ q\dot{u}_2  F^{(q)}\,.
\label{relation:FFpFq}
\ena
If $F=0$ and $F^{(p)}=0$ (or $F^{(q)}=0$), we obtain $\dot{u}_2  F^{(q)}=0$
(or $\dot{u}_1  F^{(p)}=0$), since $F=0$ is a constraint equation and
its time derivative also vanishes.
However, if we have only $F^{(p)}=0$ and $F^{(q)}=0$, $F=0$ is not
necessarily obeyed. As mentioned above, this is because $F=0$ is a
constraint equation, while $F^{(p)}=0$ and $F^{(q)}=0$ are dynamical
equations. To satisfy the constraint equation, we must impose
the initial condition. Consequently, we can solve the two
equations $F=0$ and $F^{(p)}=0$ (or $F^{(q)}=0$) instead of trying to
solve all three equations.

\newcommand{\NP}[1]{{\it Nucl.\ Phys.}\ B\ {\bf #1}}
\newcommand{\PL}[1]{{\it Phys.\ Lett.}\ B\ {\bf #1}}
\newcommand{\CQG}[1]{{\it Class.\ Quant.\ Grav.}\ {\bf #1}}
\newcommand{\CMP}[1]{{\it Comm.\ Math.\ Phys.}\ {\bf #1}}
\newcommand{\IJMP}[1]{{\it Int.\ Jour.\ Mod.\ Phys.}\ A {\bf #1}}
\newcommand{\JHEP}[1]{{\it JHEP}\ {\bf #1}}
\newcommand{\PR}[1]{{\it Phys.\ Rev.}\ D\ {\bf #1}}
\newcommand{\PRL}[1]{{\it Phys.\ Rev.\ Lett.}\ {\bf #1}}
\newcommand{\PRE}[1]{{\it Phys.\ Rep.}\ {\bf #1}}
\newcommand{\PTP}[1]{{\it Prog.\ Theor.\ Phys.}\ {\bf #1}}
\newcommand{\PTPS}[1]{{\it Prog.\ Theor.\ Phys.\ Suppl.}\ {\bf #1}}
\newcommand{\MPL}[1]{{\it Mod.\ Phys.\ Lett.}\ {\bf #1}}
\newcommand{\JP}[1]{{\it Jour.\ Phys.}\ {\bf #1}}

\end{document}